\documentclass[prx, aps, twocolumn, superscriptaddress, reprint, nobalancelastpage, nofootinbib]{revtex4-2}

\usepackage{amsmath}
\usepackage{xcolor}
\usepackage{amssymb}
\usepackage{mathtools}
\usepackage{braket}
\usepackage{txfonts}
\usepackage{comment}
\usepackage[pdftex]{graphicx}
\usepackage{bbold}
\usepackage{glossaries}

\usepackage[ruled,vlined]{algorithm2e}
\graphicspath{{figures/}}

\usepackage{xcolor}

\usepackage[colorlinks,urlcolor=blue,linkcolor=blue,citecolor=blue]{hyperref}
\usepackage{cleveref}

\DeclarePairedDelimiter{\abs}{\lvert}{\rvert}

\begin{document}

\begin{abstract}
A major challenge for the realisation of useful universal quantum computers is achieving high fidelity two-qubit entangling gate operations. However, calibration errors can affect the quantum gate operations and limit their fidelity. To reduce such errors it is desirable to have an analytical understanding and quantitative predictions of the effects that miscalibrations of gate parameters have on the gate performance. In this work, we study a systematic perturbative expansion in miscalibrated parameters of the M\o lmer-S\o rensen entangling gate, which is widely used in trapped ion quantum processors. Our analytical treatment particularly focuses on systematic center line detuning miscalibrations. Via a unitary Magnus expansion, we compute the gate evolution operator which allows us to obtain relevant key properties such as relative phases, electronic populations, quantum state purity and fidelities. These quantities, subsequently, are used to assess the performance of the gate using the fidelity of entangled states as performance metric. We verify the predictions from our model by benchmarking them against measurements in a trapped-ion quantum processor. The method and the results presented here can help design and calibrate high-fidelity gate operations of large-scale quantum computers.
\end{abstract}

\title{Analytical and experimental study of center line miscalibrations in M\o lmer-S\o rensen gates}
\author{Fernando Martínez-García}
\email{f.martinez-garcia.974203@swansea.ac.uk}
\affiliation{Department of Physics, Swansea University, Singleton Park, Swansea SA2 8PP, United Kingdom}
\author{Lukas Gerster}
\affiliation{Institut f\"ur Experimentalphysik, Universit\"at Innsbruck, Technikerstraße 25/4, 6020 Innsbruck, Austria}
\author{Davide Vodola}
\affiliation{Dipartimento di Fisica e Astronomia ``Augusto Righi", Via Irnerio 46, Bologna, Italy}
\affiliation{INFN, Sezione di Bologna, I-40127 Bologna, Italy}
\author{Pavel~Hrmo} 
\thanks{Now at: Trapped Ion Quantum Information Group, Institute for Quantum Electronics, ETH Zurich, 8093 Zurich, Switzerland}
\affiliation{Institut f\"ur Experimentalphysik, Universit\"at Innsbruck, Technikerstraße 25/4, 6020 Innsbruck, Austria}

\author{Thomas Monz}
\affiliation{Institut f\"ur Experimentalphysik, Universit\"at Innsbruck, Technikerstraße 25/4, 6020 Innsbruck, Austria}
\affiliation{AQT, Technikerstraße 17, 6020 Innsbruck, Austria}
\author{Philipp Schindler}
\affiliation{Institut f\"ur Experimentalphysik, Universit\"at Innsbruck, Technikerstraße 25/4, 6020 Innsbruck, Austria}
\author{Markus Müller}
\affiliation{Institute for Quantum Information, RWTH Aachen University, Otto-Blumenthal-Strasse 20, D-52074 Aachen Germany}
\affiliation{Peter Gr\"unberg Institute, Theoretical Nanoelectronics, Forschungszentrum J\"ulich, D-52425 J\"ulich, Germany}

\maketitle

\section{Introduction}

The implementation of a quantum information processor requires accurate initialisation, manipulation, and measurement of its qubits. Here, achieving high-fidelity single-qubit and multi-qubit entangling gates has been at the focus of intense efforts in recent years. Developments in various quantum computing platforms \cite{Saffman2016,Wendin2017,McArdle2020,Huang2020,Slussarenko2019,Bruzewicz2019a} have pushed the fidelities of the fundamental entangling gate operations \cite{Ballance2016,Hughes2020,Wang2020,Leung2018,Gaebler2016,Hong2020,Barends2014,Rol2019,Huang2019,Graham2019}. This has allowed increasingly more complex implementations of algorithms on near-term Noisy Intermediate-Scale Quantum (NISQ) devices \cite{preskill2018quantum, bharti2021noisy, montanaro2016quantum}, as well as progress towards the realisation of logical qubits that can be operated fault-tolerantly and in the regime of beneficial error correction~\cite{Egan2021,Ryan-Anderson2021,Chen2021,Hilder2021, abobeih2021fault, postler2021demonstration}. However, achieving these high-fidelity gates or further improving them requires the development and implementation of protocols aimed at the detection and correction of possible miscalibrations.

To this end, one can perform quantum process tomography~\cite{Riebe2006} to obtain a complete characterisation of the action introduced by a gate. However, this highly informative protocol has the drawback of scaling exponentially with the number of qubits involved in the gate~\cite{Chuang1997}. Moreover, errors in the resulting characterisation can appear due to the existence of state preparation and measurement (SPAM) errors or systematic errors \cite{Merkel2013}. These problems motivated the development of a number of alternative techniques for the characterisation of gate performances such as randomised benchmarking \cite{Knill2008,Mavadia2017}, cycle benchmarking~\cite{Erhard2019,PhysRevA.94.052325}, gate set tomography \cite{Blume-Kohout2017,Mavadia2017}, adaptive Bayesian inference protocols \cite{Pogorelov2017,Granade2017,Gerster2021},
and machine-learning methods~\cite{Greplova2017}. All of these protocols represent tools for learning about the imperfections of a quantum gate implementation. Moreover, having access to an analytical understanding of the effect that certain gate imperfections have on the gate performance is desirable during the implementation and calibration of a quantum gate operation. This systematic understanding of the gate interaction and its performance can then be used to conclude which imperfection limits the performance of the gate and by how much. Obtaining such understanding can not be done physical platform-agnostically, but requires studying the physics underlying the specific gate operation under consideration.

One of the leading quantum computing platforms are trapped ions \cite{cirac1995quantum, haffner2008quantum, brown2016co, blatt2008entangled, harty2014high, johanning2009quantum, leibfried2003quantum, Pogorelov2021}. Electronic states of trapped atomic ions allow one to encode qubits, and laser fields are used to manipulate their quantum information.
While single-qubit operations are relatively easy to model, implement, and calibrate, multi-qubit entangling gates that rely on interactions, mediated by the common vibrational modes of the trapped ion crystals, are significantly more complex. Examples of such gates include, but are not limited to, M{\o}lmer-S{\o}rensen (MS) \cite{Sorensen1999, sorensen2000entanglement}, as well as Raman and microwave gates~\cite{mintert2001ion, Gaebler2016, Ballance2016,weidt2016trapped, Kaufmann2017, sutherland2019versatile}. Consequently, the gate calibration requires adjusting an increased number of parameters whose effects on the gate action become more difficult to accurately model. 

In this work, we will focus on the study of the MS gate, which is based on the application of a bichromatic light field to perform correlated spin-flips over the set of qubits on which it acts.
The application of the MS gate depends on the calibration of different parameters, such as the gate time or the Rabi frequency of the interaction of the laser field with the ions. For these miscalibrations it is possible to obtain analytical expressions that help in understanding their effect on the performance of the gate~\cite{sorensen2000entanglement, Roos2008}. Here, miscalibrations cause in unwanted residual entanglement between the ions and their motional state at the end of the gate operation, or result in incorrect final internal states of the ions. However, there are other possible parameter miscalibrations that do not allow for an analytical derivation of the gate. This is the case of the center line detuning, which appears due to the two frequencies of the MS gate laser field not being centered around the carrier transition. While there is no analytical model for the effects of the center line detuning miscalibration, an understanding of its effects is highly desirable, as explained before, in order to determine its impact on the gate performance quantitatively, as well as for its calibration.

In this work, we will focus on the study of the effects that the above-mentioned center line detuning miscalibration has on the MS gate. In practice, this detuning arises when the bichromatic laser field produces an AC Stark shift due to interaction with off-resonant atomic levels~\cite{haffner2003precision, Kirchmair2009}. For this reason it cannot be simply calibrated by measuring the transition frequency using a single monochromatic laser field. The existence of this center line detuning affects the correct behaviour of the gate as it breaks both the resonance and the symmetry of the four two-photon resonant paths, introduced by the bichromatic laser field, on which the gate is based (see Fig.~\ref{fig:MS_levels_ideal} and the detailed discussion below). In order to derive the main effects of this miscalibration, we perform a perturbative study of the center line detuning and derive a semianalytical model based on a Magnus expansion~\cite{magnus1954exponential}. From this perturbative study we obtain a description for the effect of the center line detuned gate as a modified version of the evolution introduced by the ideal MS gate. This modified evolution acting over initial states in the computational basis has the effect, up to first order, of introducing unwanted relative phases and, up to second order, of changing the final populations as well as causing unwanted residual entanglement between the qubits and the motional states, decreasing the fidelity and purity of the final states. The analytical predictions that our model produces for these quantities can then be used to estimate the center line detuning of the gate in order to correct it, without requiring numerical calculations nor fitting procedures. Furthermore, we carry out a series of experiments in a trapped-ion quantum processor, against we benchmark our theoretical predictions, finding good quantitative agreement.

This paper is structured as follows: we begin by reviewing in Sec.~\ref{sec:ModelMS} the MS gate model, where we will introduce a series of experimentally relevant and possibly
miscalibrated control parameters, including the center line detuning. In Sec.~\ref{sec:effect_of_cl} we introduce the derivation of a Magnus expansion applied to understanding the effect of the center line detuned gate, with which we will be able to obtain the form of the final states after the application of the miscalibrated MS gate. In Sec.~\ref{sec:model_predictions} we will then use these final states in order to obtain expressions to predict quantities of interest such as populations, relative phases, fidelities, and purities, and compare these predictions with results obtained from numerical integration of the gate Hamiltonian. In Sec.~\ref{sec:experimental_results} we benchmark our theoretical predictions against experimental results. Finally, Sec.~\ref{sec:conclusion} presents conclusions and an outlook.

\section{M{\o}lmer-S{\o}rensen Gate Dynamics}
\label{sec:ModelMS}

In this section we review the physics underlying the MS gate~\cite{Sorensen1999, sorensen2000entanglement, Roos2008}. We will first outline the derivation of the ideal MS gate, i.e.~for the case in which all the parameters involved in the gate are correctly calibrated, yielding the desired action of the gate on the trapped-ion system. The gate is based on the application of a force that is dependent on the internal state of the ions, and takes advantage of a common vibrational mode shared between the trapped ions~\cite{James1997}. This force induces a periodic movement of the motional state of the ions in phase space. At the end of the gate the motion is returned to its original state, but with an accumulated relative phase in the internal states. This interaction can then be used to create entanglement between the qubits. The MS gate can be used to create entanglement between more than two ions with a single application and is independent of the initial vibrational state to first order~\cite{Sorensen1999, sorensen2000entanglement}. This last property provides the gate with a robustness when working with thermal states of the gate-mediating phonon mode, which can result from imperfect ground state cooling. After having explained the ideal MS gate, we will then explain how a finite center line detuning modifies the Hamiltonian of the gate. We will then use this resulting Hamiltonian as the starting point for our perturbative analysis of the center line detuned gate.

\subsection{Ideal MS gate Hamiltonian}
\label{sec:ModelMS_ideal}

\begin{figure}[ht]
\centering
\includegraphics[width=0.85\columnwidth]{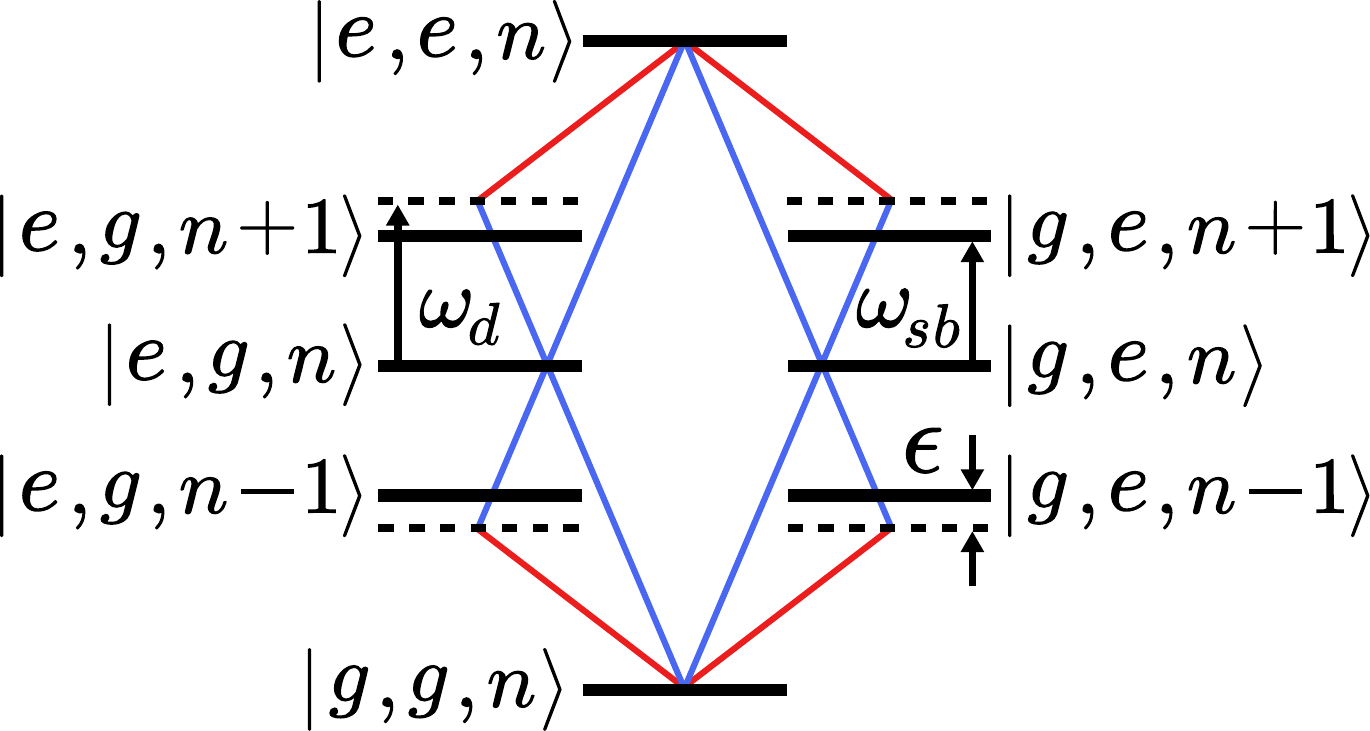}
    \caption{Energy diagram for two ions with quantised center-of-mass vibrational mode of frequency $\omega_{sb}$ interacting with a bichromatic laser which allows a resonant two-photon transition between $\ket{g,g,n}$ and $\ket{e,e,n}$. There are four different paths, each of them going through an intermediate virtual state separated by $\epsilon=\omega_{sb}-\omega_{d}$ from one of the sidebands. A similar diagram can be drawn for the two-photon transition between the states $\ket{e,g,n}$ and $\ket{g,e,n}$ introduced by the MS gate.}
    \label{fig:MS_levels_ideal}
\end{figure}

In order to derive the MS gate Hamiltonian, let us consider a system of two ions in a linear trap interacting with a bichromatic laser field of frequencies $\omega_1$ and $\omega_2$. Additionally, we will consider only the lowest frequency center-of-mass (COM) axial vibrational mode of the ions and ignore the other more energetical modes~\cite{James1997}, since the frequency difference between the COM mode and the other modes is much larger than the Rabi frequency of the driving laser field. This system may be described by the Hamiltonian
\begin{equation}
\begin{alignedat}{2}
    &H(t)= H_0&& + H_\mathrm{int}(t),\\
    &H_0 = \sum_{j=1}^2&&\frac{\omega_{eg,0}}{2}\sigma_{z,j} + \omega_{sb} (a^\dagger a + 1/2), \\
    &H_\mathrm{int}(t) = &&\sum_{j=1}^2\,\frac{\Omega(t)}{2}\left(\sigma^+_j + \sigma^-_j\right)\\& &&\cdot\Big(e^{i(\vec{k}_1\vec{x}_j-\omega_1 t+\varphi)}+e^{i(\vec{k}_2\vec{x}_j-\omega_2 t+\varphi)}+ \mathrm{H.c.}\Big),
\end{alignedat}
\end{equation}
where $\omega_{eg,0}$ is the transition frequency between the internal states $\ket{e}$ (excited) and $\ket{g}$ (ground) used to encode the qubit; $\omega_{sb}$ is the frequency of the COM mode, which defines the distance of the motional sidebands (Fig.~\ref{fig:MS_levels_ideal}) from the carrier; $a^\dagger$ and $a$ are the ladder operators related to the COM mode; $\sigma_j^+$ and $\sigma_j^-$ are the ladder operators acting on the internal states of the $j$\textsuperscript{th} ion and, similarly, $\sigma_{x,j}$, $\sigma_{y,j}$ and $\sigma_{z,j}$ represent the Pauli operators acting on the internal state of that ion; $\varphi$ is the phase of the two laser tones of frequencies $\omega_1$ and $\omega_2$ - we will consider $\varphi$ to be equal for both; $\vec{k}_1$ and $\vec{k}_2$ are the wavevectors of each laser tone; $\Omega(t)$ is the Rabi frequency, which is assumed to be equal for all ions and for both components of the laser field, and can be time-dependent for a general pulse-shape of the laser, $f(t)$,
\begin{equation}
    \Omega(t)=\Omega\, f(t).
\end{equation}

In order to implement the MS gate, the frequencies of the bichromatic laser field must be centered around the carrier transition frequency and close to the sideband transition frequency, that is, $\omega_1=\omega_{eg,0}+\omega_{d}$ and $\omega_2=\omega_{eg,0}-\omega_{d}$, with $\omega_{d}$ being a value close but not equal to $\omega_{sb}$, see Fig.~\ref{fig:MS_levels_ideal}. Introducing these values into the interaction term  of the Hamiltonian we obtain
 \begin{align}
    H_\mathrm{int}(t) =& \sum_{j=1}^2\frac{\Omega(t)}{2}\Big(e^{i(\vec{k}_1\vec{x}_j-(\omega_{eg,0}+\omega_{d}) t+\varphi)}\\&+e^{i(\vec{k}_2\vec{x}_j-(\omega_{eg,0}-\omega_{d}) t+\varphi)}+ h.c.\Big)\left(\sigma^+_j + \sigma^-_j\right).\nonumber
\end{align}

We can write $\vec{k}_i \vec{x}=\eta_{i}(a^\dagger+a)$ with $i=1,2$, where $\eta_i$ is the Lamb-Dicke parameter~\cite{leibfried2003quantum}, and since $\omega_{d}\ll \omega_{eg,0}$ we can assume $\eta_1, \eta_2 \approx\eta$. We can simplify this Hamiltonian by assuming that we are in the Lamb-Dicke regime, $\eta\sqrt{n}\ll 1$, with $n$ being the phonon number of the COM motional state, transforming to the interaction picture defined by the evolution operator generated by the free Hamiltonian, $H_0$, and introducing the sideband detuning as $\epsilon=\omega_{sb}-\omega_{d}$. As a result, we obtain the following Hamiltonian
\begin{equation}
    \label{eq:MS_ideal_H}
    \hat{H}(t)=-\eta\,\Omega(t)\,\Big(a^\dagger e^{i\epsilon t}+ae^{-i\epsilon t}\Big)\,S_\varphi,
\end{equation}
where we have applied the rotating wave approximation to keep only the terms rotating with $\epsilon$ and ignore the other fast-rotating terms that oscillate with $\omega_{sb}+\omega_{d}$ or $\omega_{eg,0}$, and we defined
\begin{equation}
    S_\varphi=S_y\cos(\varphi)+S_x\sin(\varphi),
\end{equation}
and
\begin{equation}
    S_\alpha=\frac{1}{2}\sum_{j=1}^2\sigma_{\alpha,j}, \quad \alpha=x,y,z.
\end{equation}

The Hamiltonian in Eq.~\eqref{eq:MS_ideal_H} can be integrated to obtain the corresponding evolution operator~\cite{sorensen2000entanglement, Roos2008}
\begin{equation}
\label{eq:MS_evolution_operator}
\begin{aligned}
    \hat{U}_0(t)=&D\left(\int_0^t \gamma(t')dt'\right)\\&\cdot\exp\left(i \,\mathrm{Im} \int_0^t \gamma(t')dt'\int_0^{t'} \gamma^*(t'')dt''\right),
\end{aligned}
\end{equation}
with $D(\alpha)$ being the displacement operator, $D(\alpha)=\exp(\alpha a^\dagger - \alpha^* a)$, and
\begin{equation}
    \gamma(t)= i \eta \Omega(t) e^{i\epsilon t} S_\varphi.
\end{equation}

\begin{figure}[h]
\centering
\includegraphics[width=0.95\columnwidth]{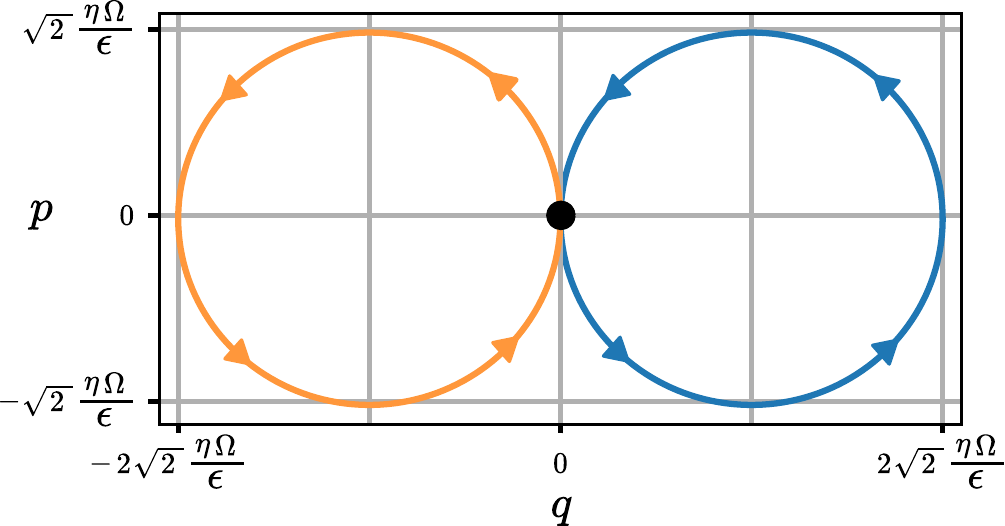}
    \caption{Trajectory in phase space induced by the gate with constant Rabi frequency with the ions in the $+1$ eigenstate of $S_\varphi$ (in orange) and the $-1$ eigenstate (in blue), with the black circle indicating the initial position. If we apply the gate for the correct amount of time the trajectory returns to the initial point in phase space. The rotation introduced by the gate is given by the area enclosed by the trajectory.}
    \label{fig:MS_loops_ideal}
\end{figure}

The parameters of the gate can be tuned in order to obtain the desired evolution
\begin{equation}
\label{eq:MS_operator}
    \mathrm{MS}_\varphi(\theta)=\exp\left(i \theta S^2_\varphi\right).
\end{equation}
One typically aims to obtain the case with $\theta=\pi/2$, for which the action of the gate on a state in the computational basis produces a maximally entangled state. In the following, we focus on the case of a time-independent laser pulse for simplicity. We note, however, that similar expressions can be derived for MS gate realizations based on time-dependent pulse shapes, which have been implemented e.g.~in~\cite{Roos2008}. Here, the evolution operator is given by
\begin{equation}
\label{eq:MS_evol_const}
\begin{aligned}
\hat{U}_0(t)=D\,&\Bigg(\frac{\eta\Omega}{\epsilon}\,(e^{i\epsilon t}-1)\,S_\varphi\Bigg)\\&\cdot\exp\left\{i\left[\frac{(\eta\Omega)^2}{\epsilon}t-\left(\frac{\eta\Omega}{\epsilon}\right)^2\sin(\epsilon\,t)\right]S^2_\varphi\right\}.
\end{aligned}
\end{equation}
We can see from Eq.~\eqref{eq:MS_evol_const} that if the total gate time $t_g$ and the sideband detuning $\epsilon$ satisfy 
\begin{equation}
\label{eq:displacement_condition}
    t_g\, \abs{\epsilon}=2\pi,
\end{equation}
then the displacement operator reduces to the identity operator. This can be understood as the gate introducing a displacement in phase space which returns to the initial state at the end of the gate after completing a loop (see Fig.~\ref{fig:MS_loops_ideal}), regardless of the initial motional and electronic state. This ensures that the internal state of the ions and their motional state decouple at the end of the gate, leaving no residual entanglement between them. Finally, if the condition in Eq.~\eqref{eq:displacement_condition} is satisfied, then one can also choose the gate parameters so they satisfy
\begin{equation}
\label{eq:angle_condition}
    \frac{(\eta\Omega)^2}{\epsilon}t_g=\frac{\pi}{2}.
\end{equation}
One can see that if these conditions are fulfilled the evolution operator in Eq.~\eqref{eq:MS_evol_const} takes the form of the maximally entangling MS gate, $\mathrm{MS}_\varphi(\pi/2)$.

While this would be the effect of the ideal MS gate, obtained by a perfect calibration of the parameters, it is important to understand what are the effects that a wrong calibration would introduce in the gate. One can see, for example, that if the relation in Eq.~\eqref{eq:angle_condition} is not satisfied, the entanglement between the internal state of the ions introduced by the gate would not be the desired one. Additionally, a miscalibration in the relation between the gate time and the sideband detuning shown in Eq.~\eqref{eq:displacement_condition} can lead to the argument of the displacement operator not being zero, leading to both a change in the motional state, and an imperfect decoupling between the motional and internal states of the ions. This can be understood as the gate causing a loop in phase space that does not return the motional state to the initial one at the end of the gate. On top of this, this miscalibration would also introduce an error in the value of $\theta$. Therefore, both of these miscalibrations introduce unwanted effects that reduce the fidelity of the gate.

The previous examples of the effects that some miscalibrations have on the gate can be easily understood thanks to having access to a closed form of the time evolution operator describing the gate. However, this will not be the case for a center line detuned gate. In the following we introduce the sources of center line detunings and show how this miscalibration affects the Hamiltonian of the MS gate.

\subsection{Center line detuned MS gate}

In the following we will consider that a center line detuning can appear due to two different contributions. First, we consider that the transition frequency between $\ket{g}$ and $\ket{e}$ can be affected by an AC-Stark shift $\lambda_{AC}(t)\propto \Omega(t)^2$ appearing due to the interaction of the laser with off-resonant atomic levels \cite{haffner2003precision,Kirchmair2009}, which will be time-dependent due to the pulse-shape of the laser, $f(t)$. Due to this, the transition frequency will have the form
\begin{equation}
    \omega_{eg}(t)=\omega_{eg,0}+\lambda_{AC}\, f(t)^2.
\end{equation}
The other contribution to the center line detuning that we consider is due to a shift, $\lambda_l$, of the bichromatic laser frequencies. Considering this miscalibration, the frequencies are given by
\begin{equation}
    \begin{aligned}
    \omega_1=\omega_{eg,0}+\omega_{d}+\lambda_l,\\
    \omega_2=\omega_{eg,0}-\omega_{d}+\lambda_l.
    \end{aligned}
\end{equation}

\begin{figure}[h]
\centering
\includegraphics[width=0.85\columnwidth]{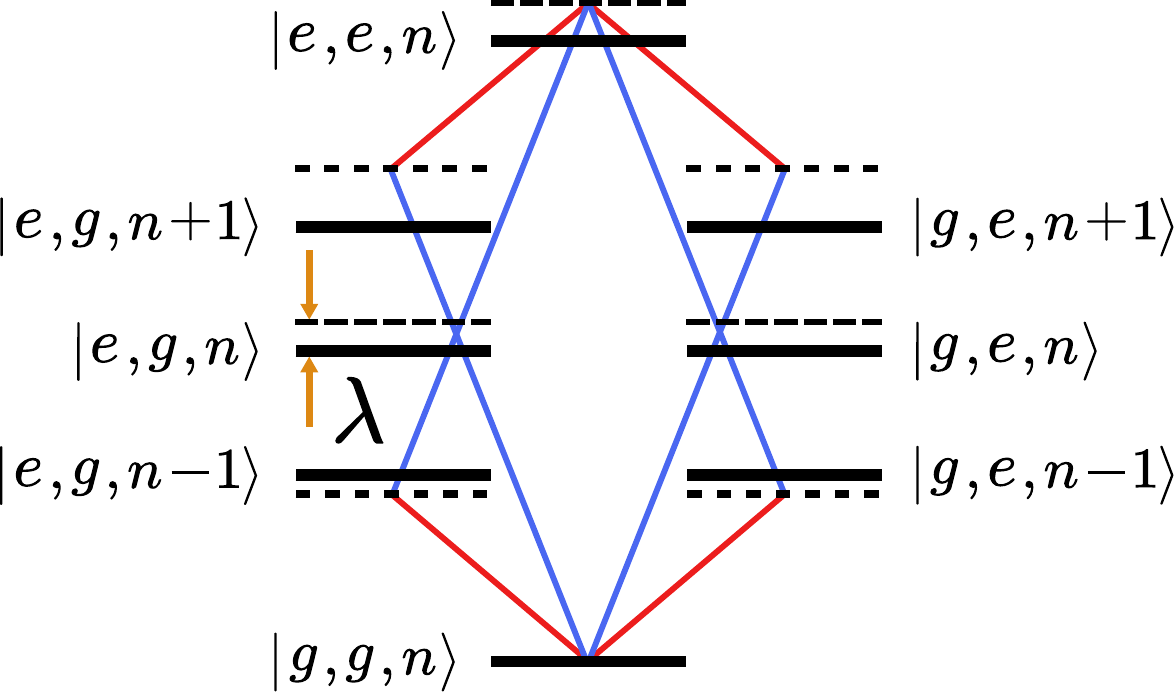}
    \caption{Miscalibration of the MS gate due to the existence of a center line detuning, denoted by $\lambda$. This miscalibration breaks both the symmetry of the four paths in which the MS gate is based and the two-photon resonance. As a consequence, this miscalibration reduces the fidelity of the gate.}
    \label{fig:MS_levels_cl}
\end{figure}

Both of these contributions cause a detuning of the mean value of the bichromatic frequencies from the carrier transition frequency, given by
\begin{equation}
\label{eq:time_dep_cl}
    \lambda(t) = \lambda_{AC}\, f(t)^2 - \lambda_l.
\end{equation}
The effect of this miscalibration on the level structure of the MS gate at a given time is shown in Fig.~\ref{fig:MS_levels_cl}.

\begin{figure}[ht]
\centering
\includegraphics[width=0.95\columnwidth]{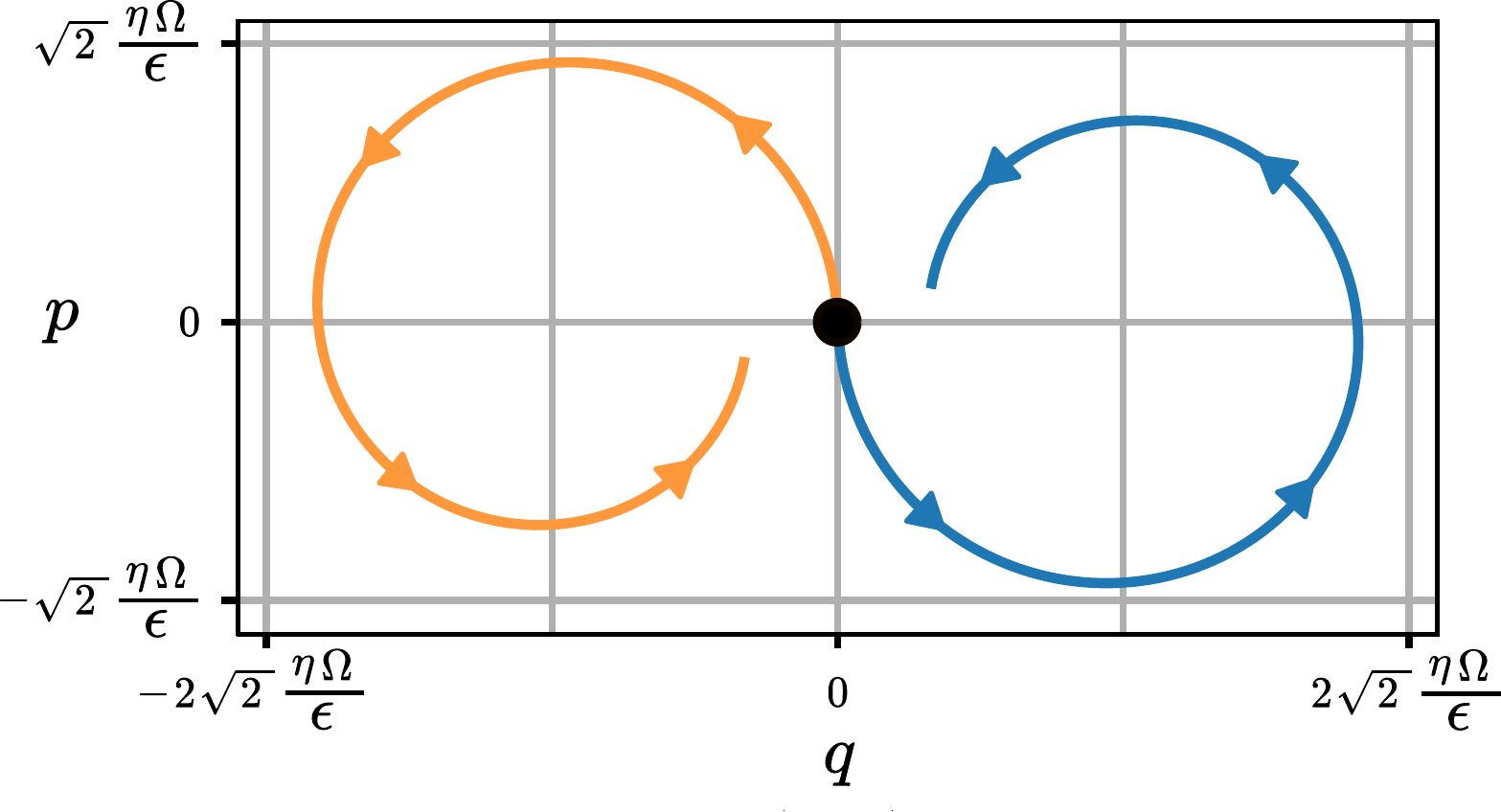}
    \caption{Trajectory in phase space induced by the gate with a constant laser pulse ($f(t)=1$ for the duration of the gate) for the cases where the ions are in the $+1$ eigenstate of $S_\varphi$ (in orange) and the $-1$ eigenstate (in blue), and the black circle indicating the initial position, with a center line detuning of $\lambda(t)/\epsilon=\lambda/\epsilon=0.1$. The center line detuning miscalibration deforms the loop as compared to the ideal case in Fig.~\ref{fig:MS_loops_ideal}, causing the final motional state to be different from the initial one, therefore introducing and entanglement between the motional states and the ionic states. Additionally, the trajectory in this case is dependent on the initial motional state, with the case shown being for initial $\ket{n=0}$. This effect also causes the final internal state to be different from the ideal one (not shown).}
    \label{fig:MS_loops_cl}
\end{figure}

We can introduce these changes into the model and follow a similar derivation as in Sec.~\ref{sec:ModelMS_ideal}. In this case, instead of changing into the interaction picture defined by the evolution operator generated by $H_0$, we transform to an interaction picture defined by
\begin{equation}
    V(t)=\exp\left[i\left(\sum_{j=1}^2\frac{\omega_{eg,0}-\lambda_{l}}{2}\sigma_{z,j}+\omega_{sb}(a^\dagger a +1/2)\right)t\right].
\end{equation}
The resulting Hamiltonian is given by
\begin{align}
    \label{eq:interaction_H_time_dep}
    \hat{H}(t)=\lambda(t)\,S_z-\eta\Omega(t)&(a^\dagger e^{i\epsilon t}+a e^{-i\epsilon t})\,S_\varphi.
\end{align}
Therefore, the effect of the center line detuning miscalibrations is the appearance of an unwanted $S_z$ term in the Hamiltonian. This unwanted term causes the resulting time evolution to differ from the ideal gate, as illustrated in Fig.~\ref{fig:MS_loops_cl}. It also causes the miscalibrated gate to evade an analytical closed-form solution.

In the following section we present a perturbative study based on a Magnus expansion, to analyse the effect of the center line detuning on gate performance.

\section{Perturbative study of the center line detuned MS gate}
\label{sec:effect_of_cl}

As we saw in the previous section, the Hamiltonian including the center line detuning has the form of the ideal Hamiltonian of the MS gate plus an additional term accounting for the center line detuning miscalibration. In the following we will rescale the time as $\tau = \epsilon\,t$, which is dimensionless since the center-line detuning $\epsilon$ is a frequency. This rescaling is convenient since it makes our study independent of specific choices for the gate parameters. The time-dependent Schrödinger equation becomes
\begin{equation}
    i \epsilon \frac{d}{d\tau}\phi(\tau)= \hat{H}(\tau/\epsilon)\phi(\tau).
\end{equation}
The time evolution is governed by the rescaled Hamiltonian, $\hat{\mathcal{H}}(\tau)$, given by
\begin{equation}
\label{eq:time_dep_cl_H}
    \hat{\mathcal{H}}(\tau)=\hat{\mathcal{H}}_\textrm{ideal}(\tau)+\tilde{\mathcal{H}}_\textrm{cl}(\tau),
\end{equation}
where
\begin{equation}
\begin{aligned}
	\hat{\mathcal{H}}_\textrm{ideal}(\tau)&=-\tilde{\Omega}(\tau/\epsilon)(a^\dagger e^{i\tau}+ae^{-i\tau})\,S_\varphi, \\
	\hat{\mathcal{H}}_\textrm{cl}(\tau)&=\tilde\lambda(\tau/\epsilon)\,S_z,
\end{aligned}
\end{equation}
with $\tilde{\Omega}(\tau/\epsilon)=\eta\Omega(\tau/\epsilon)/\epsilon$ and $\tilde{\lambda}(\tau/\epsilon)=\lambda(\tau/\epsilon)/\epsilon$. In order to simplify the analysis, in the following, we will consider a square pulse-shape. In this case we can write $\tilde{\lambda}(\tau/\epsilon)=\tilde\lambda=(\lambda_{AC}-\lambda_l)/\epsilon$ and $\tilde{\Omega}(\tau/\epsilon)=\tilde\Omega=\eta\Omega/\epsilon$, which satisfies $\tilde{\Omega}=1/2$ when we consider the conditions in Eqs.~\eqref{eq:displacement_condition} and~\eqref{eq:angle_condition}. We note that for the study of a given time-dependent pulse shape, one can follow a similar derivation as the one that we will show, at the cost of introducing the pulse shape in the respective numerical integrals that will appear.

Our goal thus consists of obtaining the evolution operator, $\hat{\mathcal{U}}(\tau)$, associated to the Hamiltonian in Eq.~\eqref{eq:time_dep_cl_H}. In this section we will study the effect of the center line detuning miscalibration by using a Magnus expansion approach where we consider this unwanted term as a perturbation. This perturbative study is justified since typical values of the center line detuning miscalibration represent only a small fraction of the value of the sideband detuning. Without loss of generality, we will continue to assume that the common phase of both laser tones is $\varphi=0$.

\subsection{Magnus expansion}

Since we already know the form of the evolution introduced by the unperturbed Hamiltonian $\hat{\mathcal{H}}_\textrm{ideal}(\tau)$, given by the rescaled version of Eq.~\eqref{eq:MS_evolution_operator}
\begin{equation}
    \hat{\mathcal{U}}_0(\tau)=D\Big[F(\tau)\,S_y\Big]\cdot\exp\left[i G(\tau) S^2_y\right],
\end{equation}
where $F(\tau) = \tilde{\Omega}\,(e^{i\tau}-1)$ and $G(\tau) = \tilde{\Omega}^2\Big(\tau-\sin(\tau)\Big)$, we can transform to another rotating frame defined by this free evolution operator, $\hat{\mathcal{U}}_0(\tau)$. In this frame, the Hamiltonian describing the evolution of the system is
\begin{equation}
\tilde{\mathcal{H}}(\tau)=\hat{\mathcal{U}}^\dagger_0(\tau)\hat{\mathcal{H}}_\textrm{cl}\hat{\mathcal{U}}_0(\tau).
\end{equation}
In this rotating frame, we separate the perturbation term due to the center line detuning of the initial Hamiltonian, $\hat{\mathcal{H}}(\tau)$, from the one that allows for an analytical solution. From now on, we will focus on studying the term arising from the center line detuning as a perturbation to the Hamiltonian of the ideal gate (with the perturbative parameter being $\tilde{\lambda}$) by using a Magnus expansion~\cite{magnus1954exponential, Blanes2010}, which has the advantage of producing a unitary perturbative evolution operator at any order. With this approach, we can represent the operator that describes the evolution due to this Hamiltonian, $\tilde{\mathcal{U}}(\tau)$, in the following exponential form:
\begin{equation}
	\tilde{\mathcal{U}}(\tau)=\exp\Big[M(\tau)\Big],
\end{equation}
where the Magnus exponent, $M(\tau)$, is given by the following series
\begin{equation}
M(\tau)=\sum_{j=1}M_j(\tau),
\end{equation}
with the first terms being
\begin{align}
	&M_1(\tau)=-i\int_0^\tau \tilde{\mathcal{H}}(t_1)dt_1,\label{eq:M1}\\
	&M_2(\tau)=-\frac{1}{2}\int_0^\tau \int_0^{t_1} [\tilde{\mathcal{H}}(t_1),\tilde{\mathcal{H}}(t_2)]dt_2 dt_1.\label{eq:M2}
\end{align}
Here, $M_1(\tau)$ and $M_2(\tau)$ have a first and second order dependence on $\tilde{\lambda}$, respectively. Thus, considering only these terms and ignoring other possible ones which have a higher order dependence on the center line detuning, the evolution operator can be approximated by
\begin{equation}
\begin{aligned}
	\tilde{\mathcal{U}}^{(2)}(\tau)\approx \mathbb{1}+ M_1(\tau)+\Big(M_2(\tau)+M^2_1(\tau)/2\Big).
\end{aligned}
\end{equation}
We can now use this result with the free evolution operator to obtain an approximation up to second order in Magnus expansion, $\hat{\mathcal{U}}^{(2)}(\tau)$, of the evolution, $\hat{\mathcal{U}}(\tau)$, introduced by the Hamiltonian in Eq.~\eqref{eq:time_dep_cl_H}
\begin{equation}
\label{eq:msm_magnus_evolution}
\hat{\mathcal{U}}^{(2)}(\tau)\approx \hat{\mathcal{U}}_0(\tau)+ \hat{\mathcal{U}}_0(\tau)M_1(\tau)+\hat{\mathcal{U}}_0(\tau)\Big(M_2(\tau)+M^2_1(\tau)/2\Big).
\end{equation}
One can use this second order evolution operator or, more generally, the corresponding evolution operator obtained by a perturbation expansion up to $K$\textsuperscript{th} order, to obtain the action that the gate has over an initial state after a normalised time $\tau_g$. In the following we will consider the initial states $\ket{\sigma,\sigma',n}$ in the computational basis where $\sigma, \sigma' \in \{ g,e \}$, and study the action of the center line detuned gate on them in a sum over $k$\textsuperscript{th} order perturbative state corrections $\ket{\psi^{(k)}_{\sigma,\sigma',n}}$ with coefficients $\tilde{\lambda}^k$.   
The resulting state $\ket{\Psi^{(K)}_{\sigma,\sigma',n}}$ can be written as:
\begin{equation}
\label{eq:nth_order_final_state}
    \ket{\Psi^{(K)}_{\sigma,\sigma',n}}=\hat{\mathcal{U}}^{(K)}(\tau_g)\ket{\sigma,\sigma',n}=\ket{\psi_{\sigma,\sigma',n}}-\sum_{k=1}^K\tilde{\lambda}^k\ket{\psi^{(k)}_{\sigma,\sigma',n}},
\end{equation}
where $\ket{\psi_{\sigma,\sigma',n}}$ is the target state of the ideal gate.

We will begin by studying the first order Magnus expansion, from which we will be able to write the perturbed evolution as a unitary operator acting on the qubit space. From this we will see that the only linear effect that the center line detuning has on the gate is the appearance of an unwanted relative phase. Studying the second order Magnus expansion will allow us to capture more exactly the dependencies of other quantities, namely the populations, fidelities, and purities, on the center line detuning.

\subsection{First order Magnus expansion}
\label{sec:first_order}

In order to study the first order dependence of the MS gate on the center line detuning, we consider only the first two terms in Eq.~\eqref{eq:msm_magnus_evolution}
\begin{equation}
    \hat{\mathcal{U}}^{(1)}(\tau)\approx \hat{\mathcal{U}}_0(\tau)+ \hat{\mathcal{U}}_0(\tau)M_1(\tau).
\end{equation}
In the absence of any center line detuning, the final state after the application of the ideal MS gate for a normalised gate time $\tau_g=2\pi$ (see Eq.~\eqref{eq:displacement_condition}) for each initial state are given by
\begin{align}
    &\ket{\psi_{g,g,n}}=\hat{\mathcal{U}}_0(\tau_g)\ket{g,g,n}=\frac{e^{i\pi/4}}{\sqrt{2}}\Big(\ket{g,g,n} - i\ket{e,e,n}\Big),\label{eq:ideal_gg}\\
    &\ket{\psi_{e,e,n}}=\hat{\mathcal{U}}_0(\tau_g)\ket{e,e,n}=\frac{e^{i\pi/4}}{\sqrt{2}}\Big(-i\ket{g,g,n} + \ket{e,e,n}\Big),\\
    &\ket{\psi_{g,e,n}}=\hat{\mathcal{U}}_0(\tau_g)\ket{g,e,n}=\frac{e^{i\pi/4}}{\sqrt{2}}\Big(\ket{g,e,n} + i\ket{e,g,n}\Big),\\
    &\ket{\psi_{e,g,n}}=\hat{\mathcal{U}}_0(\tau_g)\ket{e,g,n}=\frac{e^{i\pi/4}}{\sqrt{2}}\Big(i\ket{g,e,n} +\ket{e,g,n}\Big).
\end{align}
One can then calculate the first order corrections to the final states (see Appendix~\ref{app:numerical_coefficients}), given by
\begin{equation}
\begin{aligned}
    &\ket{\psi^{(1)}_{g,g,n}}=\sum_{m\ge 0}\Big[i f^\textrm{odd}_{n,m}I^m_n\left(\ket{e,g,m}+\ket{g,e,m}\right)\label{eq:gg_first_correction}\\&+f^\textrm{even}_{n,m}\left(\left(I^n_m+I^m_n\right)\ket{g,g,m}+\left(I^n_m-I^m_n\right)\ket{e,e,m}\right)\Big],
\end{aligned}
\end{equation}
\begin{equation}
\begin{aligned}
    &\ket{\psi^{(1)}_{e,e,n}}=\sum_{m\ge 0}\Big[-i f^\textrm{odd}_{n,m}I^m_n\left(\ket{e,g,m}+\ket{g,e,m}\right)\\&+f^\textrm{even}_{n,m}\left(\left(I^n_m+I^m_n\right)\ket{g,g,m}+\left(I^n_m-I^m_n\right)\ket{e,e,m}\right)\Big],
\end{aligned}
\end{equation}
\begin{equation}
\begin{aligned}
    \ket{\psi^{(1)}_{g,e,n}}=\ket{\psi^{(1)}_{e,g,n}}=i\sum_{m\ge 0}I^n_m f^\textrm{odd}_{n,m} \Big(\ket{g,g,m}+\ket{e,e,m}\Big)\label{eq:ge_first_correction}.
\end{aligned}
\end{equation}
Here, we defined
\begin{equation}
\label{eq:feven_fodd}
    f^\textrm{even}_{n,m}=\frac{(1+(-1)^{n-m})}{2}, \quad f^\textrm{odd}_{n,m}=\frac{(1-(-1)^{n-m})}{2},
\end{equation}
and $I^n_m$, which is a matrix (see Appendix~\ref{app:numerical_coefficients} and Fig.~\ref{fig:an_values}) obtained from numerical integrations of
\begin{equation}
    I^n_m=\frac{i}{2}\int_0^{\tau_g} e^{i G(\tau)}\bra{m}D[F(\tau)]\ket{n}d\tau.
\end{equation}
While these coefficients depend on the pulse-shape and on the phonon motional states, they do not have any dependence on the center-line detuning. This implies that, for obtaining corrections to the final states, we only need to calculate $I^n_m$ once for that given pulse-shape. The coefficients obtained can then be used with Eq.~\eqref{eq:nth_order_final_state} for the calculation of the final state $\ket{\Psi^{(1)}_{\sigma,\sigma',n}}$ up to first order for any value of the center line detuning. In our constant pulse-shape case, all of these coefficients have the form of a real value multiplied by $(-1+i)$. The code used to obtain these coefficients can be found in Ref.~\cite{numericalCode}, which also includes the numerical calculation of the second order coefficients appearing in Sec.~\ref{sec:second_order}.

\begin{figure}[ht]
	\centering
	\includegraphics[width=0.9\columnwidth]{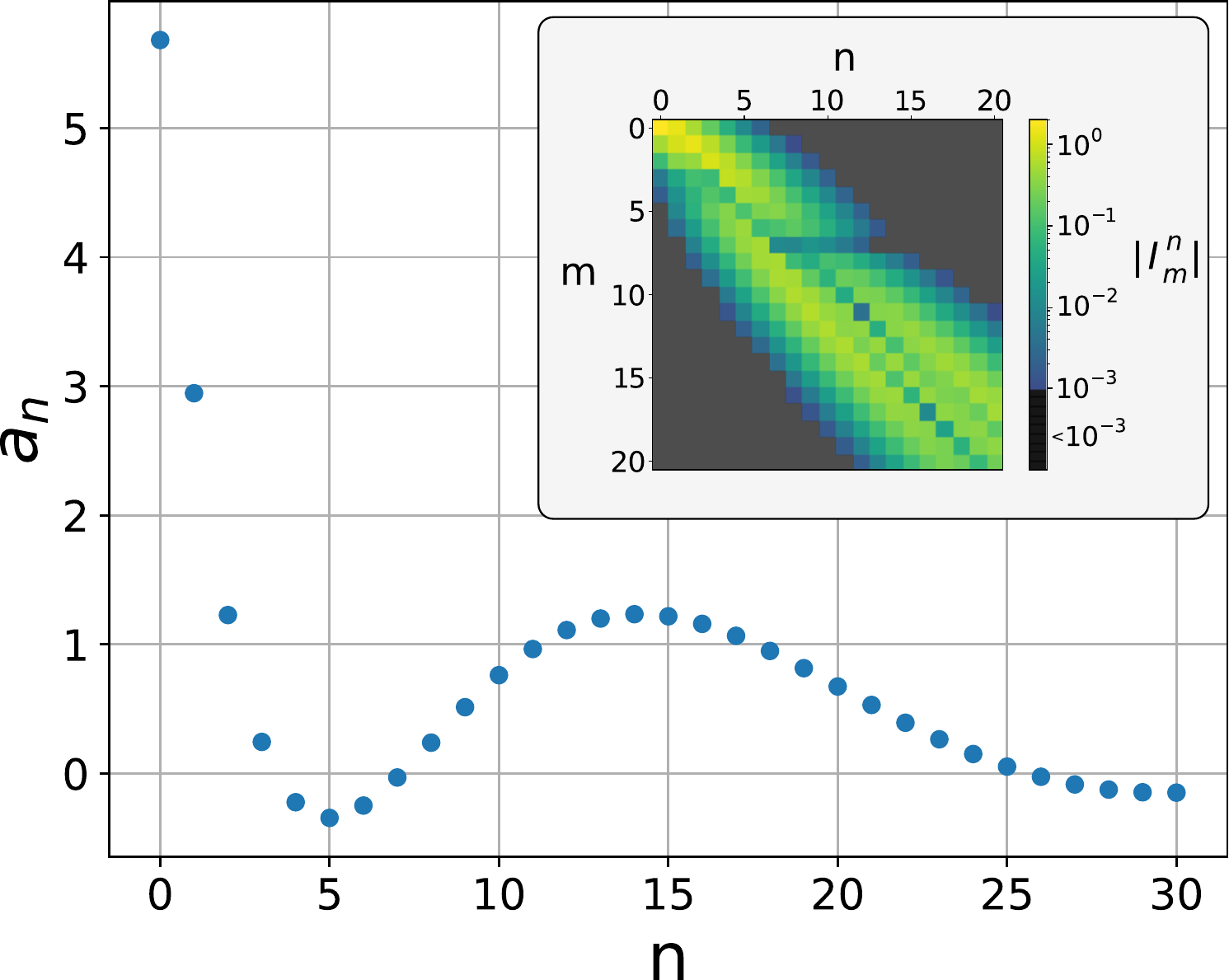}
	\caption{Values of the numerical coefficients $a_n$, defined in Eq.~\eqref{eq:an_def}. These values appear in the predictions of our model due to the first order corrections to the final state. Therefore, for the first values of $n$, the first order effects decrease as $n$ increases (see, for example, Eq.~\eqref{eq:detuned_to_ideal}, Fig.~\ref{fig:bloch}, and Fig.~\ref{fig:phases_comparison}). These low values of $n$ are the most relevant in the experiment since the initial motional state of the ions is cooled before the application of the gate. An inset is shown for the absolute values of the $I^n_m$ coefficients, from which the $a_n$ coefficients are obtained by using Eq.~\eqref{eq:an_def}.}
	\label{fig:an_values}
\end{figure}

We now show that up to first order in $\tilde{\lambda}$, the center line detuning does not introduce any unwanted entanglement between the internal and motional states. This can be seen by computing the density matrix of the state $\ket{\Psi^{(1)}_{\sigma,\sigma',n}}$ that from Eq.~\eqref{eq:nth_order_final_state} takes the form:
\begin{equation}
\ket{\Psi^{(1)}_{\sigma,\sigma',n}} \bra{\Psi^{(1)}_{\sigma,\sigma',n}} = \ket{\Psi^{(1), tr}_{\sigma,\sigma',n}} \bra{\Psi^{(1), tr}_{\sigma,\sigma',n}} \otimes \ket{n} \bra{n} + O(\tilde{\lambda}^2),
\end{equation}
where the states $\ket{\Psi^{(1), tr}_{\sigma,\sigma',n}}$ are the following
\begin{align}
    \ket{\Psi^{(1), tr}_{g,g,n}}&=\frac{1}{\sqrt{2}}\Big[(1-i a_n \tilde\lambda )\ket{g,g}-i\ket{e,e}\Big]\label{eq:gg_state_first},\\
    \ket{\Psi^{(1), tr}_{e,e,n}}&=\frac{1}{\sqrt{2}}\Big[-i\ket{g,g}+(1+i a_n\tilde{\lambda})\ket{e,e}\Big],\\
    \ket{\Psi^{(1),tr}_{g,e,n}}&=\frac{1}{\sqrt{2}}\Big(\ket{g,e} + i\ket{e,g}\Big),\\
    \ket{\Psi^{(1),tr}_{e,g,n}}&=\frac{1}{\sqrt{2}}\Big(\ket{e,g} + i\ket{g,e}\Big)
\end{align}
and we defined the real numbers $a_n$ as
\begin{equation}
\label{eq:an_def}
    a_n=4I^n_n/(-1+i).
\end{equation}
A representation of the values $a_n$ for different values of $n$ is shown in Fig.~\ref{fig:an_values}.

\begin{figure}[t]
	\centering
	\includegraphics[width=0.8\columnwidth]{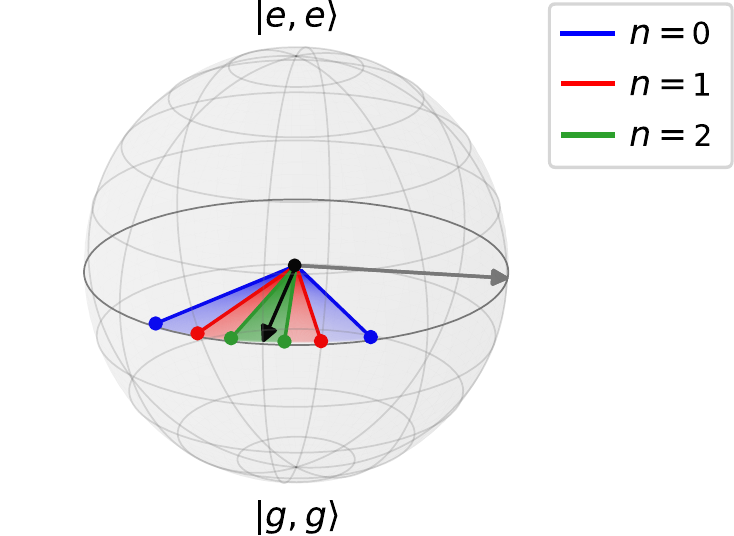}
	\caption{Representation in the Bloch sphere spanned by $\ket{g,g}$ and $\ket{e,e}$ of the effect on the initial state $\ket{g,g,n}$ of the center line detuned gate as obtained from the first order Magnus expansion as in Eq.~\eqref{eq:gg_state_first} for $n=0,1,2$. The black arrow indicates the ideal final state $(\ket{g,g}-i\ket{e,e})/\sqrt{2}$ and the grey arrow indicates the state $(\ket{g,g}+\ket{e,e})/\sqrt{2}$. These states show a relative phase different from the ideal one, $(\ket{g,g}+\exp(i\phi^{(1)}_{g,g,n}(\tilde{\lambda}))\ket{e,e})/\sqrt{2}$, given by Eq.~\eqref{eq:phase_expansion} and Eq.~\eqref{eq:phase_first} in Sec.~\ref{sec:relative_phase}. For a given initial Fock state we represent the position of the final state, up to first order approximation, with the extremal dots of the corresponding color representing the case with $\tilde\lambda=-0.1$ (left dot) and $\tilde\lambda=0.1$ (right dot). Since the first order expansion gives a linear behaviour in the center line detuning, the position of the final state for an intermediate value of $\tilde\lambda$ will be comprised between these two extremal points.}
	\label{fig:bloch}
\end{figure}

From these results one can obtain that the action of the center line detuned MS gate over the qubits is, up to first order, a unitary operator given by
\begin{equation}
\label{eq:first_unitary}
    \hat{\mathcal{U}}^{(1),tr}_n(\tau_g)=\frac{1}{\sqrt{2}}\begin{pmatrix}
    1-i a_n\tilde\lambda  & 0 & 0 & -i\\
    0 & 1 & i & 0\\
    0 & i & 1 & 0\\
    -i & 0 & 0 & 1+i a_n\tilde\lambda
\end{pmatrix}.
\end{equation}
Therefore, the center line detuning does not introduce, up to first order, any unwanted entanglement between the internal and motional states. In order to compare this unitary with the one from the ideal MS gate we consider that $1-i a_n \tilde\lambda \approx\exp(-i\tilde\lambda a_n)$, from which we can identify
\begin{equation}
\label{eq:detuned_to_ideal}
    \hat{\mathcal{U}}^{(1),tr}_n(\tau_g)=R_z(-a_n\tilde\lambda)\hat{\mathcal{U}}_0(\tau_g)R_z(-a_n\tilde\lambda),
\end{equation}
where
\begin{equation}
    R_z(\phi)=\exp\left(i\,\phi\,S_z/2\right).
\end{equation}
This shows that the predominant effect to first order, of the center line detuning is an unwanted relative phase shift for the initial states $\ket{e,e,n}$ and $\ket{g,g,n}$ (see Fig.~\ref{fig:bloch}). Other relevant effects, e.g.~in the final electronic and motional populations, quantum state fidelities or purity of the final qubit states appear in higher order, and thus require an expansion at least to second order in the center line detuning.

\begin{figure*}[ht]
	\centering
	\includegraphics[width=\textwidth]{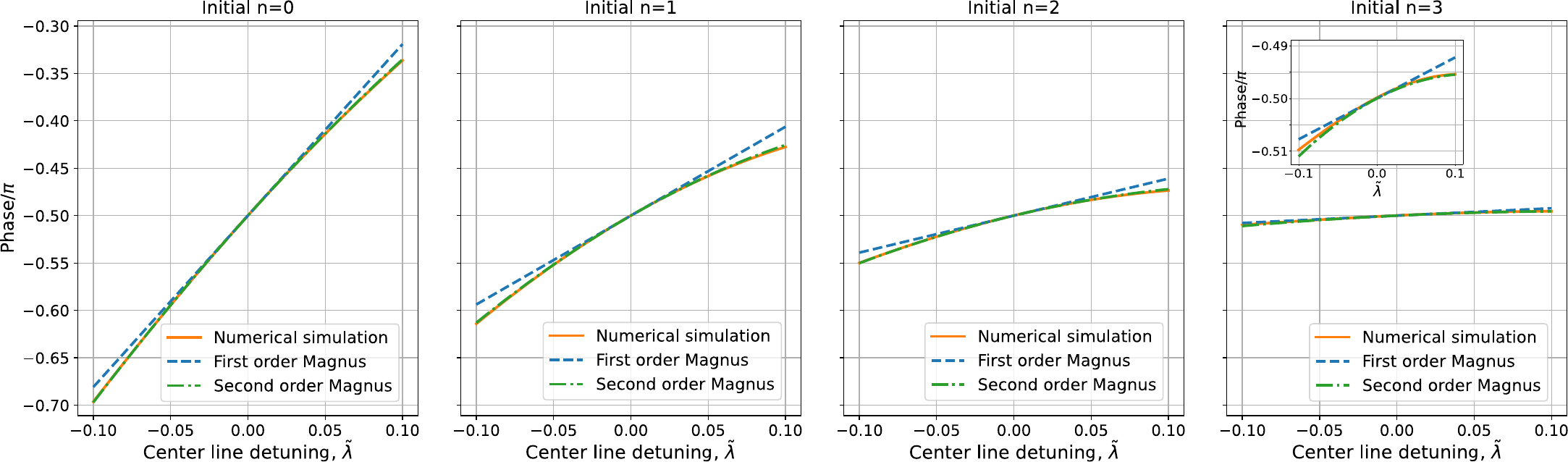}
	\caption{Relative phases after the detuned MS gate with initial state $\ket{g,g,n}$ (in this example we consider initial COM phonon numbers $n=0,1,2,3$) for the numerical integration of the Hamiltonian in Eq.~\eqref{eq:time_dep_cl_H} and for the results from the first and second order Magnus expansion, $\phi^{(1)}_{g,g,n}$ and $\phi^{(2)}_{g,g,n}$, obtained by using Eq.~\eqref{eq:phase_expansion}. The ideal relative phase is $-\pi/2$ since the ideal final state is $(\ket{g,g}-i\ket{e,e})/\sqrt{2}$. The second order terms improve the phase estimation as compared to the first order case. Very minor differences with respect to the numerics are expected and result from not accounting for higher order corrections. We note that for the cases shown, the dependence of the phase error on the center line detuning decreases as $n$ increases, in accordance to what is shown in Fig.~\ref{fig:bloch}. An inset is included for the case with initial $n=3$ for clarity.}
	\label{fig:phases_comparison}
\end{figure*}

\subsection{Second order Magnus expansion}
\label{sec:second_order}

In order to obtain the final states up to second order, we have to calculate the terms corresponding to that order, as given by Eq.~\eqref{eq:nth_order_final_state}. These are given by (see Appendix~\ref{app:numerical_coefficients})
\begin{align}
    &\ket{\psi^{(2)}_{g,g,n}}=\sum_{m\ge 0} J^n_{+,m}\ket{g,g,m}-J^n_{-,m}\ket{e,e,m},\label{eq:gg_second_correction}\\
    &\ket{\psi^{(2)}_{e,e,n}}=\sum_{m\ge 0} -J^n_{-,m}\ket{g,g,m}+J^n_{+,m}\ket{e,e,m},\\
    &\ket{\psi^{(2)}_{g,e,n}}=\ket{\psi^{(2)}_{e,g,n}}=\sum_{m\ge 0}\left(J^n_{1,m}-J^n_{2,m}\right)\Big(\ket{g,e,m}+\ket{e,g,m}\Big)\label{eq:ge_second_correction}.
\end{align}
with $J^n_{+,m}$, $J^n_{-,m}$, $J^n_{1,m}$ and $J^n_{2,m}$ being coefficients obtained numerically (see Appendix~\ref{app:numerical_coefficients}). Similarly as the $I^n_m$ coefficients, these coefficients do not depend on the center line detuning. They depend on the pulse-shape, need to be calculated only once and can then be used to calculate the final state up to second order, $\ket{\Psi^{(2)}_{\sigma, \sigma', n}}$, for any value of the center line detuning.

In this second order Magnus expansion, the effect of the gate can no longer be expressed as a unitary operator acting on the qubit states. This is because the states obtained from second (or higher) order have a residual entanglement between internal and motional states. As a consequence, the states obtained after tracing the phonons are no longer pure states. However, this second order expansion will allow us to obtain a better approximation of the final state, as we will see in the following section.

\section{Predictions of the model}
\label{sec:model_predictions}

\begin{figure*}[ht]
	\centering
	\includegraphics[width=\textwidth]{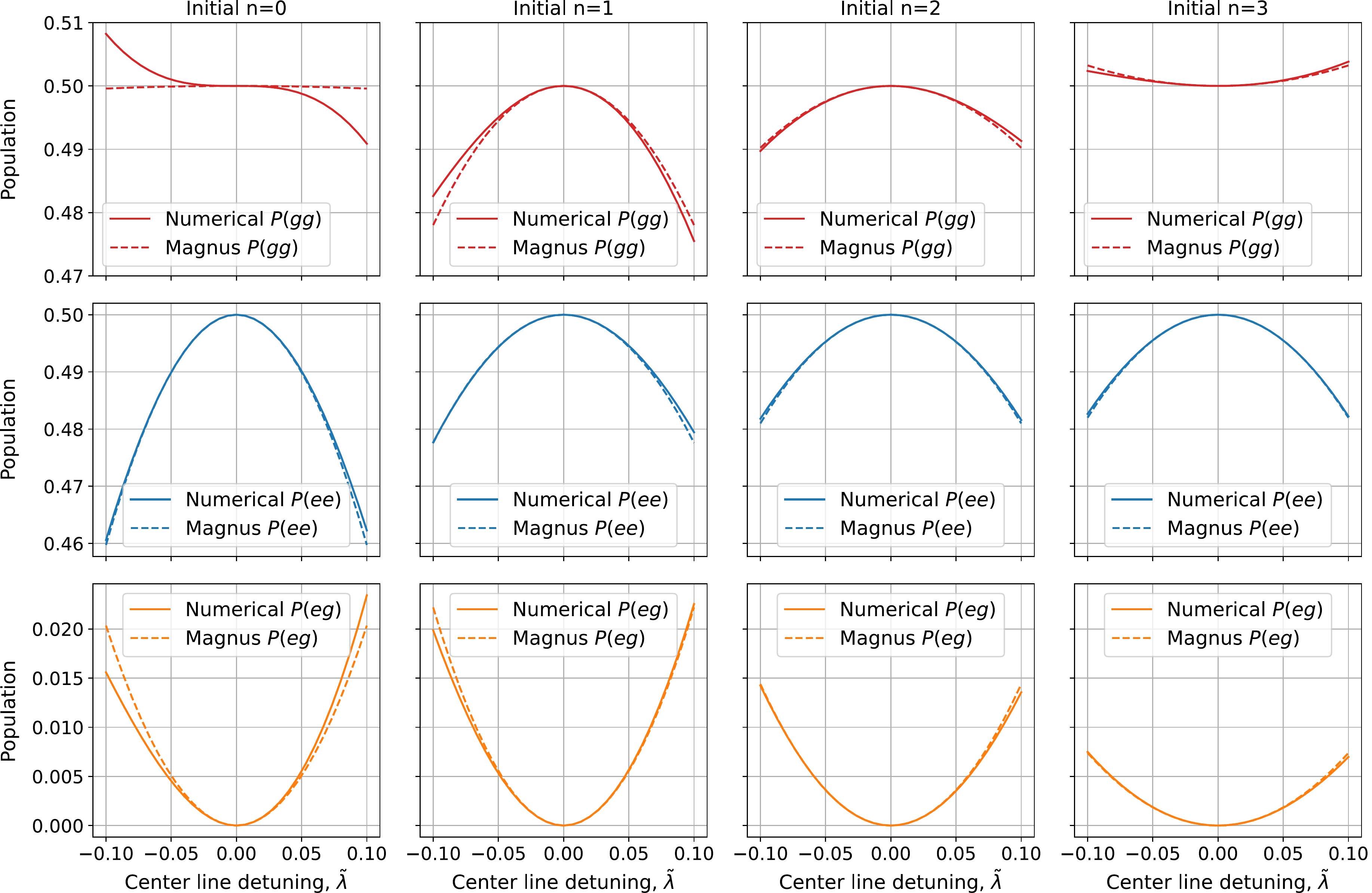}
	\caption{Populations after the action of the detuned MS gate over the initial state $\ket{g,g,n}$ (in this example we consider the cases with $n=0,1,2,3$) obtained by numerical integration of the Hamiltonian in Eq.~\eqref{eq:time_dep_cl_H} and with the expressions for the second order Magnus expansion shown in Eq.~\eqref{eq:pop_gg}-\eqref{eq:pop_eg}. None of these populations have a linear dependence with the center line detuning, which justifies the use of the second order Magnus expansion for their study. While the quadratic terms captures the behaviour of the populations with the center line detuning, one can also see that the model has slight deviations from the numerics in higher order terms. This is specially the case for the behaviour of $P(gg)$ for the initial state $\ket{g,g,0}$, in which the second order term has almost no importance, making the third order term (which is not considered in our calculations) dominant.}
	\label{fig:populations_comparison}
\end{figure*}

Having derived the final states after the application of the center line detuned gate by using a Magnus expansion, we now estimate the effect that the center line detuning miscalibration has on the phase (Sec.~\ref{sec:relative_phase}), populations (Sec.~\ref{sec:populations}), fidelities (Sec.~\ref{sec:fidelity}), and purities (Sec.~\ref{sec:purity}) of the final state with respect to the ideal one. While the following study can be performed for any initial state by using the results shown in Sec.~\ref{sec:effect_of_cl}, in the following we will focus on the case of $\ket{g,g,n}$ as initial state. We will also discuss how these results can be generalised for the more experimentally relevant case where ions are in thermal motional states.

\subsection{Phase error}
\label{sec:relative_phase}

Following our perturbative approach, we can write the final relative phase of the state in Eq.~\eqref{eq:nth_order_final_state} when considering the initial state $\ket{\sigma,\sigma',n}$ as the relative phase of the target state, $\phi^{(0)}_{\sigma,\sigma',n}$, plus the terms related to the $k$\textsuperscript{th} order correction, up to the considered $K$\textsuperscript{th} order:
\begin{equation}
\label{eq:phase_expansion}
    \phi^{(K)}_{\sigma,\sigma',n}(\tilde{\lambda})=\phi^{(0)}_{\sigma,\sigma',n}+\sum_{k=1}^K \tilde\lambda^k \delta \phi^{(k)}_{\sigma,\sigma',n}
\end{equation}

Looking at the results from the first order Magnus expansion, we can see that the predominant effects of this miscalibration is over the relative phase of the final internal states. From Eq.~\eqref{eq:first_unitary} and Eq.~\eqref{eq:detuned_to_ideal} we can see that the initial states $\ket{g,e,n}$ and $\ket{e,g,n}$ are not affected by a center line detuning in first order. As for the initial state $\ket{g,g,n}$, it has a first order correction given by (see Eq.~\eqref{eq:detuned_to_ideal})
\begin{equation}
\label{eq:phase_first}
    \delta \phi^{(1)}_{g,g,n}=a_n.
\end{equation}
This causes the final state to have an error in the final relative phase as compared to the target state in Eq.~\eqref{eq:gg_state_first}, of relative phase $\phi^{(0)}_{g,g,n}=-\pi/2$.

For the case of the initial state $\ket{e,e,n}$ one can obtain, after following the same derivation, that the relative phase introduced by the center line detuning is the same but of opposite sign. Therefore, the relative phase has a leading first order perturbative term dependent on the center line detuning. A visual representation of this effect is shown in Fig.~\ref{fig:bloch}.

Additionally, one can use the results from the second order Magnus expansion to improve the relative phase prediction. In order to do this, one can, for the initial state $\ket{g,g,n}$ case for example, calculate the coherence element between $\ket{e,e}$ and $\ket{g,g}$ of the final state, after tracing over the phonon mode, 
\begin{equation}
\label{eq:second_order_coherence}
    \rho^{ee,gg\,(2)}_{g,g,n}(\tilde\lambda)=\frac{-i+\tilde{\lambda}a_n+\tilde{\lambda}^2b_n}{2}, 
\end{equation}
where
\begin{equation}
    b_n = -(1-i)\Big((J^n_{+,n})^*-J^n_{-,n}\Big)+\sum_{m\neq n}\Big(I^n_m-I^m_n\Big)\Big(I^n_m+I^m_n\Big)^* f^\textrm{even}_{n,m}.
\end{equation}
Calculating the argument of this coherence term up to second order, we obtain
\begin{equation}
\label{eq:phase_second}
    \delta \phi^{(2)}_{g,g,n}=\textrm{Re}(b_n),
\end{equation}
which can be used with Eq.~\eqref{eq:phase_expansion} to obtain the second order correction.
A comparison between the relative phases predicted up to first and second order, $\phi^{(1)}_{g,g,n}$ and $\phi^{(2)}_{g,g,n}$, and the one obtained from the numerical integration of the Hamiltonian in Eq.~\eqref{eq:time_dep_cl_H}, for different initial motional states is shown in Fig.~\ref{fig:phases_comparison}.

This result can be generalised for the case where the initial motional state is in a mixed state defined by a thermal distribution with a mean number of phonons $\bar{n}$, $p_{\bar{n}}(n)$, given by
\begin{equation}
    p_{\bar{n}}(n)=\frac{\bar{n}^n}{(\bar{n}+1)^{n+1}}
\end{equation}
In this case, the relative phases in first, $\phi^{(1)}_{g,g,\bar{n}}$, and second order, $\phi^{(2)}_{g,g,\bar{n}}$, are given by
\begin{align}
    &\phi^{(K)}_{g,g,\bar{n}}=\sum_{n=0} p_{\bar{n}}(n)\, \phi^{(K)}_{g,g,n},\quad K=1,2.
\end{align}
Since each of the populations of the thermal state is affected by a different error in the final relative phase, one can expect that the center line detuning also introduces a dephasing of the final internal state superposition, causing decoherence in the quantum states.

\subsection{Populations}
\label{sec:populations}

From the final states obtained from the first order Magnus expansion in Sec.~\ref{sec:first_order} one can obtain that the center line detuning does not introduce a first order correction of the populations when considering the elements of the computational basis as initial states. Thus, we have to consider the states obtained from the second order Magnus expansion. The populations obtained from $\ket{\Psi^{(2)}_{g,g,n}}$ are given by
\begin{align}
    &P^{(2)}_{g,g,n}(gg,\tilde{\lambda})=\frac{1}{2}+c_{g,g,n}\tilde\lambda^2, \label{eq:pop_gg}\\
    &P^{(2)}_{g,g,n}(ee,\tilde{\lambda})=\frac{1}{2}+c_{e,e,n}\tilde\lambda^2,\label{eq:pop_ee}\\
    &P^{(2)}_{g,g,n}(ge,\tilde{\lambda})=P_{g,g,n}(eg,\tilde{\lambda})=c_{e,g,n}\tilde\lambda^2\label{eq:pop_eg}.
\end{align}
The expressions for $c_{g,g,n}$, $c_{e,g,n}$ and $c_{e,e,n}$ can be found in Appendix~\ref{app:numerical_coefficients}. From these equations one can see that the final populations for this case have no linear dependence with the center line detuning. The center line detuning does not only introduce an error in the populations of $\ket{g,g}$ and $\ket{e,e}$, but also leads to population of states $\ket{e,g}$ and $\ket{g,e}$, which are ideally unpopulated when considering the initial $\ket{g,g,n}$ state. A comparison between these predictions for the populations and the numerics is shown in Fig.~\ref{fig:populations_comparison}. In this figure one can see that our model correctly predicts the behaviour obtained from the numerics, with differences arising from third-order terms, which we are not considering.

\subsection{Fidelity}
\label{sec:fidelity}

Here, we will study the fidelity $F_{g,g,n}(\tilde\lambda)$ of the final internal state when applying a center line detuned MS gate to $\ket{g,g,n}$ compared to the ideal final internal state
\begin{equation}
\label{eq:ideal_traced}
    \ket{\psi^{tr}_{g,g,n}}=\frac{1}{\sqrt{2}}\Big(\ket{g,g} - i\ket{e,e}\Big).
\end{equation}
This fidelity will be given by
\begin{equation}
    F_{g,g,n}(\tilde\lambda)=\bra{\psi^{tr}_{g,g,n}}\rho_{g,g,n}(\tilde\lambda)\ket{\psi^{tr}_{g,g,n}},
\end{equation}
where $\rho_{g,g,n}(\tilde\lambda)$ is the density matrix of the internal state after applying the center line detuned MS gate.
In a similar way as for the populations studied in the previous section, this fidelity has no linear dependence with the center line detuning. Due to this, we consider the fidelity up to second order Magnus expansion, for which we will need to consider the second order density matrix, which has the form
\begin{equation}
\small
    \rho^{(2)}_{g,g,n}(\tilde\lambda)=\begin{pmatrix}
    P^{(2)}_{g,g,n}(gg,\tilde{\lambda}) & 0 & 0 & \rho^{ee,gg\,(2)}_{g,g,n}(\tilde\lambda)\\
    0 & P^{(2)}_{g,g,n}(eg,\tilde{\lambda}) & \rho^{ge,eg\,(2)}_{g,g,n}(\tilde\lambda) & 0\\
    0 & \rho^{eg,ge\,(2)}_{g,g,n}(\tilde\lambda) & P^{(2)}_{g,g,n}(ge,\tilde{\lambda}) & 0\\
    \rho^{gg,ee\,(2)}_{g,g,n}(\tilde\lambda) & 0 & 0 & P^{(2)}_{g,g,n}(ee,\tilde{\lambda})\end{pmatrix},
\end{equation}
where $\rho^{gg,ee\,(2)}_{g,g,n}(\tilde\lambda)$ is the complex conjugate of $\rho^{ee,gg\,(2)}_{g,g,n}(\tilde\lambda)$ in Eq.~\eqref{eq:second_order_coherence}, and one can easily check that all the coherences of $\ket{e,g}$ and $\ket{g,e}$ with $\ket{e,e}$ and $\ket{g,g}$ are zero thanks to the appearance of products of $f^\textrm{odd}_{n,m}$ and $f^\textrm{even}_{n,m}$ defined in Eq.~\eqref{eq:feven_fodd}. Finally, the last element left, $\rho^{ge,eg\,(2)}_{g,g,n}(\tilde\lambda)$, has the following form:
\begin{equation}
    \rho^{ge,eg\,(2)}_{g,g,n}(\tilde\lambda)=\tilde{\lambda}^2 \sum_{m\ge 0} f^\textrm{odd}_{n,m}\abs{I^m_n}^2,
\end{equation}
although this element will not be needed to calculate the fidelity or the purity in Sec.~\ref{sec:purity} in second order approximation.

Using this density matrix we can obtain the fidelity up to second order, which is given by
\begin{equation}
\label{eq:fidelity_2nd}
    F^{(2)}_{g,g,n}(\tilde\lambda)=1+\frac{\tilde\lambda^2}{2}\left(c_{g,g,n} + c_{e,e,n} - \textrm{Im}(b_n)\right),
\end{equation}

\begin{figure}[h]
    \centering
    \includegraphics[width=\columnwidth]{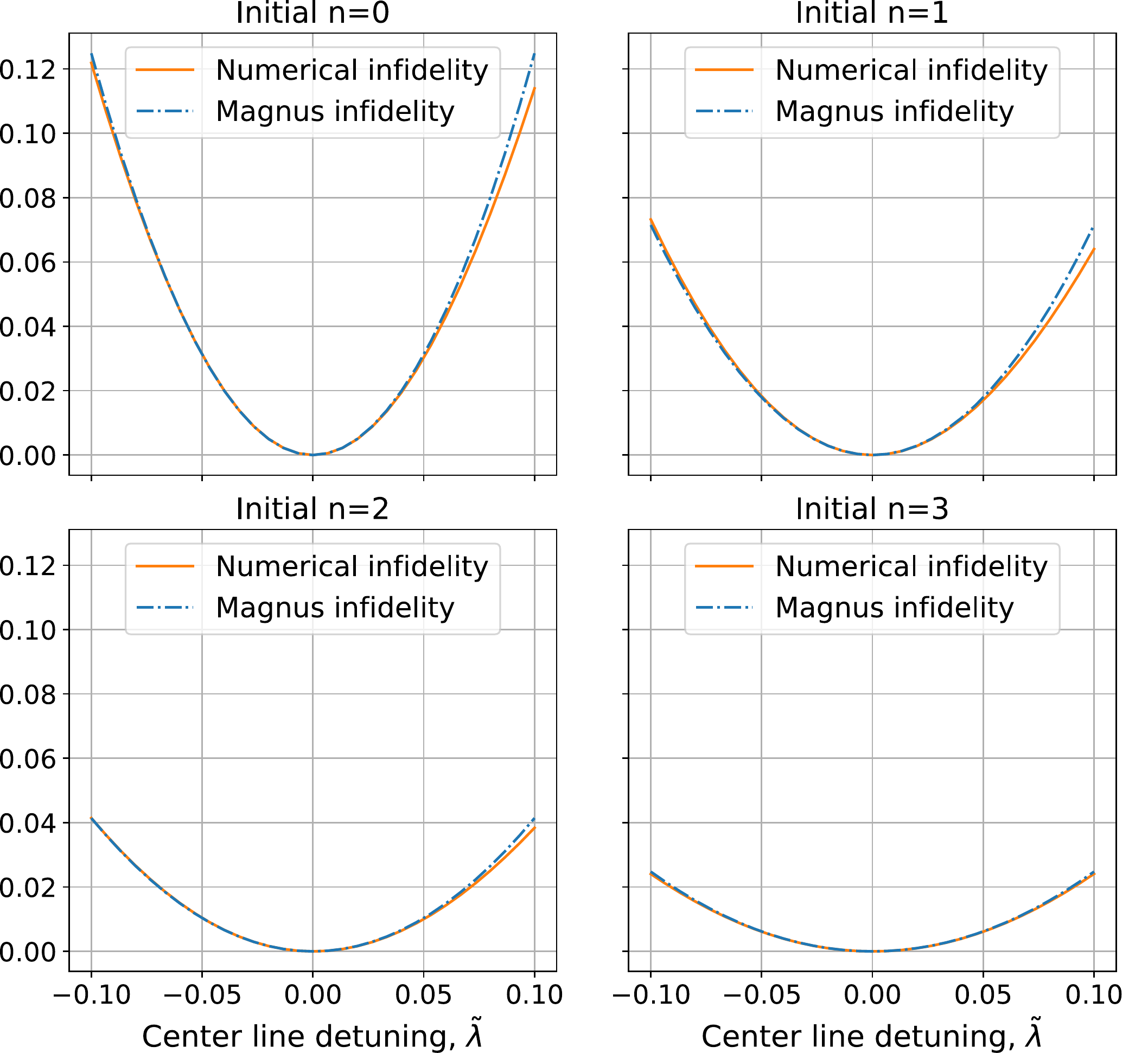}
    \caption{Infidelity of the target internal state obtained for the ideal MS gate (no center line detuning) in Eq.~\eqref{eq:ideal_traced} compared to the state obtained by numerically integrating the Hamiltonian in Eq.~\eqref{eq:time_dep_cl_H}, and the fidelity estimation obtained from the second order Magnus expansion for initial motional states with $n=0,1,2,3$. The initial state used is $\ket{g,g,n}$, with the second order Magnus expansion prediction being Eq.~\eqref{eq:fidelity_2nd}.}
    \label{fig:fidelity_paper}
\end{figure}

We compare this result for the final state fidelities with the ones obtained numerically by integrating the Hamiltonian in Eq.~\eqref{eq:time_dep_cl_H} in Fig.~\ref{fig:fidelity_paper}.

Additionally, we can generalise this result for the case of having an initial thermal state, for which we obtain
\begin{equation}
    \label{eq:fidelity_thermal_2nd}
    F^{(2)}_{g,g,\bar{n}}(\tilde\lambda)=\sum_n p_{\bar{n}}(n) F^{(2)}_{g,g,n}(\tilde\lambda).
\end{equation}

\subsection{Purity}
\label{sec:purity}

Finally, we will study how much the center line detuned MS gate transforms the initial pure state $\ket{g,g,n}$ into a mixed state. In order to quantify we will to consider again the density matrix of the final internal state, $\rho_{g,g,n}(\tilde\lambda)$, in order to obtain its purity, $\gamma_{g,g,n}(\tilde\lambda)$, given by
\begin{equation}
    \gamma_{g,g,n}(\tilde\lambda)=\mathrm{Tr}\,\Big[\rho_{g,g,n}(\tilde\lambda)^2\Big].
\end{equation}

\begin{figure}[ht]
    \centering
    \includegraphics[width=\columnwidth]{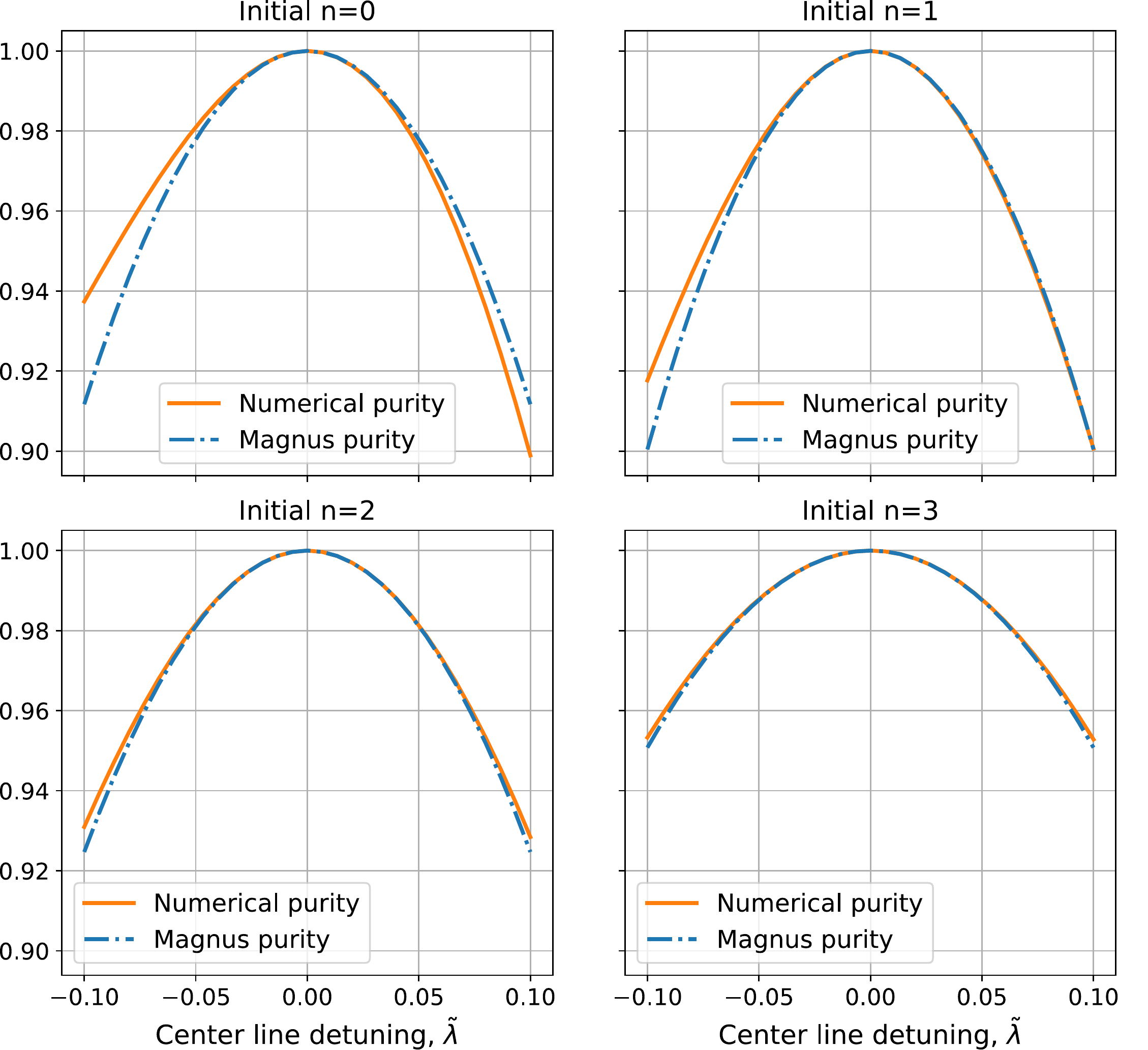}
    \caption{Purity of the final state obtained numerically by taking $\ket{g,g,n}$ as the initial state compared to the purity obtained by using Eq.~\eqref{eq:purity_second} derived from the second order Magnus expansion. The perturbative expression manages to capture the predominant second order effect of the center line detuning, with the differences from the numerics arising from terms of higher order not considered.}
    \label{fig:purity_paper}
\end{figure}

In a similar way as in the previous cases, the final purity does not show a first order dependence with the center line detuning. Therefore, we will consider the purity up to second order in the center line detuning, for which we obtain
\begin{align}
\label{eq:purity_second}
    \gamma^{(2)}_{g,g,n}(\tilde\lambda)=\mathrm{Tr}&\,\Bigg[\Big(\rho^{(2)}_{g,g,n}(\tilde\lambda)\Big)^2\Bigg]\nonumber\\&=1-\tilde{\lambda}^2\left(\textrm{Im}(b_n)-\frac{a_n^2}{2}-c_{g,g,n}-c_{e,e,n}\right).
\end{align}
From this result we can see that the phonon mixing introduced by the center line detuned MS gate affects the purity of the final state to second order in the center line detuning parameter. A comparison between the result in Eq.~\eqref{eq:purity_second} and the numerics is shown in Fig.~\ref{fig:purity_paper}.

\section{Experimental validation}
\label{sec:experimental_results}

In the previous section we compared the predictions of our model with the numerical simulation results. Now we will compare some of these model predictions with results obtained in the experiment.

\subsection{Experimental apparatus}

The following experiments are performed on $^{40}\mathrm{Ca}^+$ ions confined in a microstructured radio-frequency ion surface trap \cite{brandl2016cryogenic}. Qubits are encoded in the computational subspace formed by the $4^2S_{1/2,-1/2}\equiv \ket{g}$ electronic ground state and the metastable excited  $3^2D_{5/2,-1/2}\equiv \ket{e}$ state. We mediate the MS gate using the axial center-of-mass (COM) mode of the two-ion crystal. The entangling operations are performed using a narrow-linewidth ($<10$ Hz) diode laser at 729 nm with the two frequency tones imprinted using an acousto-optic modulator. For any practical operating conditions, the frequency difference between the COM mode and any of the other modes of an $N$ ion crystal is much larger than the Rabi frequency of the driving field \cite{James1997}, such that we can neglect the coupling to all other modes. For all experiments, the ions are initially Doppler cooled on the $4^2S_{1/2} \xrightarrow{} 4^2P_{1/2}$ transition, followed by sideband cooling of the COM mode. State readout is performed by fluorescence detection with a photomultiplier tube~\cite{schindler2013quantum}.

\subsection{Measurement protocol}

To prepare the Fock states of the COM mode, we start by preparing the ground state $\ket{g,g,0}$ using standard sideband cooling and optical pumping techniques. We then apply a $\pi$-pulse to one of the ions to spectroscopically decouple it in an auxiliary level of the $D_{5/2}$ manifold~\cite{schindler2013quantum}. We then apply alternating $\pi$-pulses on the blue and red sideband, where each pulse adds a single phonon. For odd phonon states, we apply a $\pi$-pulse on the carrier following the sideband pulses~\cite{wineland1998experimental}. Finally, we retrieve the hidden ion from the auxiliary level. For the initial ground state cooling we find a mean phonon number of $\bar{n}\approx0.05$. After the preparation sequence for Fock states $n>0$ we measure $5\%$ population outside of the target electronic state, which decreases the signal-to-noise ratio of the measurement. We use an additional repumping step to return this population to the electronic ground state, but this leaves us with a corresponding error in the initial prepared Fock state.

After preparing the desired $\ket{g,g,n}$ state, we will need to control the center line detuning of our MS gates. This center line detuning is here introduced on purpose by changing the frequencies of the laser fields by $\lambda$ from their ideal value. This causes the Hamiltonian of the MS gate to have the form
\begin{align}
    \hat{H}_\mathrm{exp}=-\eta\Omega&(a^\dagger e^{i \epsilon t}+ae^{-i \epsilon t})\\&\cdot\left[S_y\cos(\varphi+\lambda\, t)+S_x\sin(\varphi+\lambda\, t)\right]\nonumber.
\end{align}
This Hamiltonian can be obtained from the one that we considered in Eq.~\eqref{eq:interaction_H_time_dep} with a time-independent value $\lambda(t)=\lambda$ by performing a picture change defined by $V(t)=R_z(2\,\lambda\, t)$. Therefore, after taking into account this picture change, all the results of our model can be used for this experiment. Although in the experiment the laser pulse is switched on and off adiabatically using a Blackman like shape~\cite{Schindler2008}, each of this switch on and off requires only $\sim 4\%$ of the gate time. Due to this, we approximate the pulse-shape as a constant one.

The evolution introduced by the MS gate in the experiment is, up to first order, given by
\begin{equation}
    \hat{\mathcal{U}}^{(1),tr}_{n,\mathrm{exp},\varphi}(t_g, t_0)=R_z(2\lambda t_f)\hat{\mathcal{U}}^{(1),tr}_{n,\varphi}(t_g)R_z(-2\lambda t_0),
\end{equation}
where $t_0$ is the time at the beginning of the gate, $t_f=t_0+t_g$ is the time at the end, and we introduced $\varphi$ to denote the phase of the MS gate.

\begin{figure}[h]
    \centering
    \includegraphics[width=0.5\columnwidth]{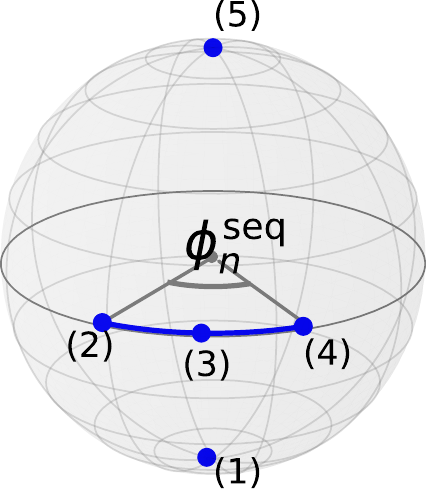}
    \caption{Representation of the effect of the MS gate sequence given by our model up to first order in the center line detuning. A center line detuned MS gate is applied to the initial state, $\ket{g,g,n}$ represented in (1). While the target state after a calibrated MS gate should be $1/\sqrt{2}(\ket{g,g,n}-i\ket{e,e,n})$ in (2), the state after this first gate differs by a phase as given by Eq.~\eqref{eq:detuned_to_ideal}, resulting in the state (3). The second detuned MS gate introduces a final phase, resulting in the state in (4), before applying the ideal entangling operation. The final state $\ket{e,e,n}$ in (5) after this sequence is obtained if the second MS gate has a relative phase with respect to the first one given by $\varphi_d=\phi^\mathrm{seq}_{n}(\lambda)$.}
    \label{fig:fidelities_LD}
\end{figure}

We perform a sequence of two center line detuned MS gates, where we consider that the second one has a relative phase of $\varphi_d$ with respect to the first one. For this sequence, the resulting population predicted by our model is, up to first order, given by
\begin{equation}
\begin{aligned}
    P(&ee, \lambda)=\abs{\bra{e,e}\hat{\mathcal{U}}^{(1),tr}_{n,\mathrm{exp},\varphi_d}(2\,t_g, t_g)\hat{\mathcal{U}}^{(1),tr}_{n,\mathrm{exp},0}(t_g, 0)\ket{g,g}}^2
    \\&=\abs{\bra{e,e}\mathrm{MS}_{\varphi_d}(\pi/2)R_z\left[-\phi^\mathrm{seq}_{n}(\lambda)\right]\mathrm{MS}_0(\pi/2)\ket{g,g}}^2
    \\&=\frac{1+\cos\Big[2\,\varphi_d + \phi^\mathrm{seq}_{n}(\lambda)\Big]}{2},
    \end{aligned}
\end{equation}
where
\begin{equation}
\label{eq:phi_sequence}
    \phi^\mathrm{seq}_{n}(\lambda) = \frac{2\,\lambda\,a_n}{\epsilon}
\end{equation}
represents the phase introduced by the center line detuning of the gates obtained from the first order terms of our model.

\begin{figure}[h]
    \centering
    \includegraphics[width=0.95\columnwidth]{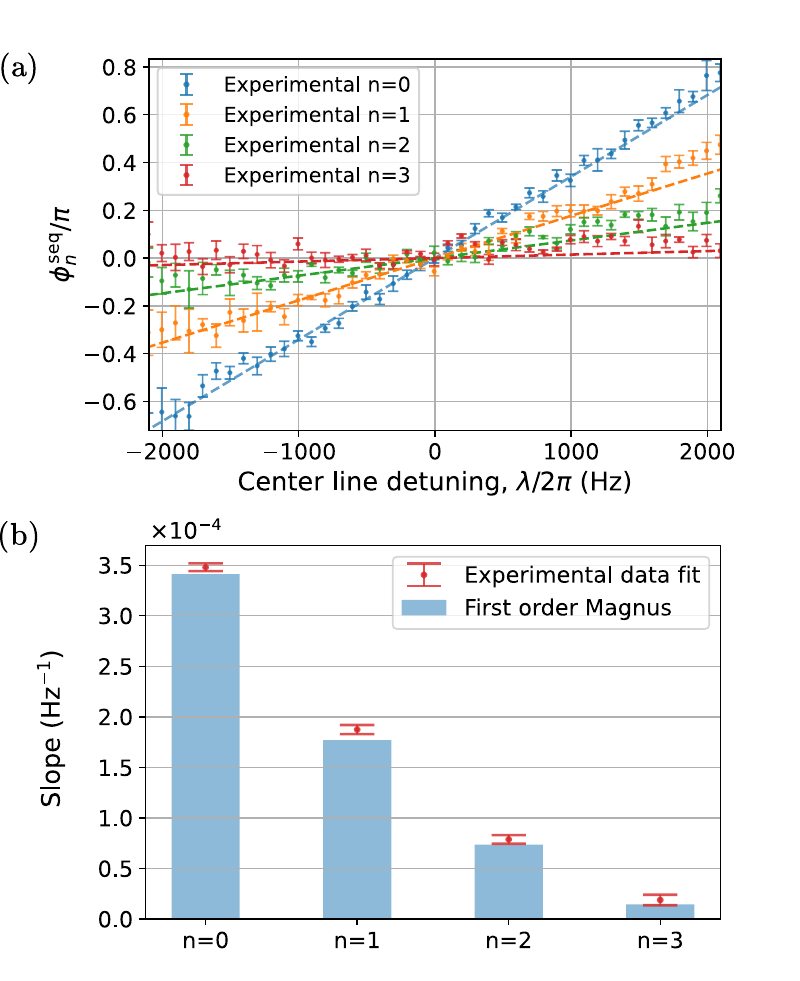}
    \caption{(a) Experimental measurements of the phase $\phi_\mathrm{seq}(\lambda)$ obtained by applying MS gates with $\epsilon=-2\pi\cdot 11$kHz to initial states $\ket{g,g,n}$ with $n=0,1,2,3$. The dashed lines represent the corresponding estimated values of our model, obtained by using Eq.~\eqref{eq:phi_sequence}. The asymmetry of the error bars stems from the asymmetric behaviour of the $P(eg)+P(ge)$ outcomes observed in both the numerical simulations shown in Fig.~\ref{fig:populations_comparison} and the experimental results in Fig.~\ref{fig:experimental_populations_all}, which causes an asymmetry on the contrast of the phase oscillation. (b) Comparison of the slope values obtained from first order Magnus expansion and from performing a linear fit using the experimental results.}
    \label{fig:experimental_results}
\end{figure}

Using sequences of this type, we can measure the final population of $\ket{e,e}$ for a given center line detuning, while varying the value of $\varphi_d$. This data can then be used to experimentally obtain the values of $\phi^\mathrm{seq}_{n}(\lambda)$ for that center line detuning by fitting a cosine to the measurement outcomes, and compare them with the predicted values from our model. The fit includes amplitude and offset as free parameters, as higher order effects, dephasing, and SPAM errors will affect the amplitude of the observed oscillations.
The comparison between the experimental results and the prediction from our model is shown in Fig.~\ref{fig:experimental_results}.

\begin{figure}[h]
    \centering
    \includegraphics[width=0.95\columnwidth]{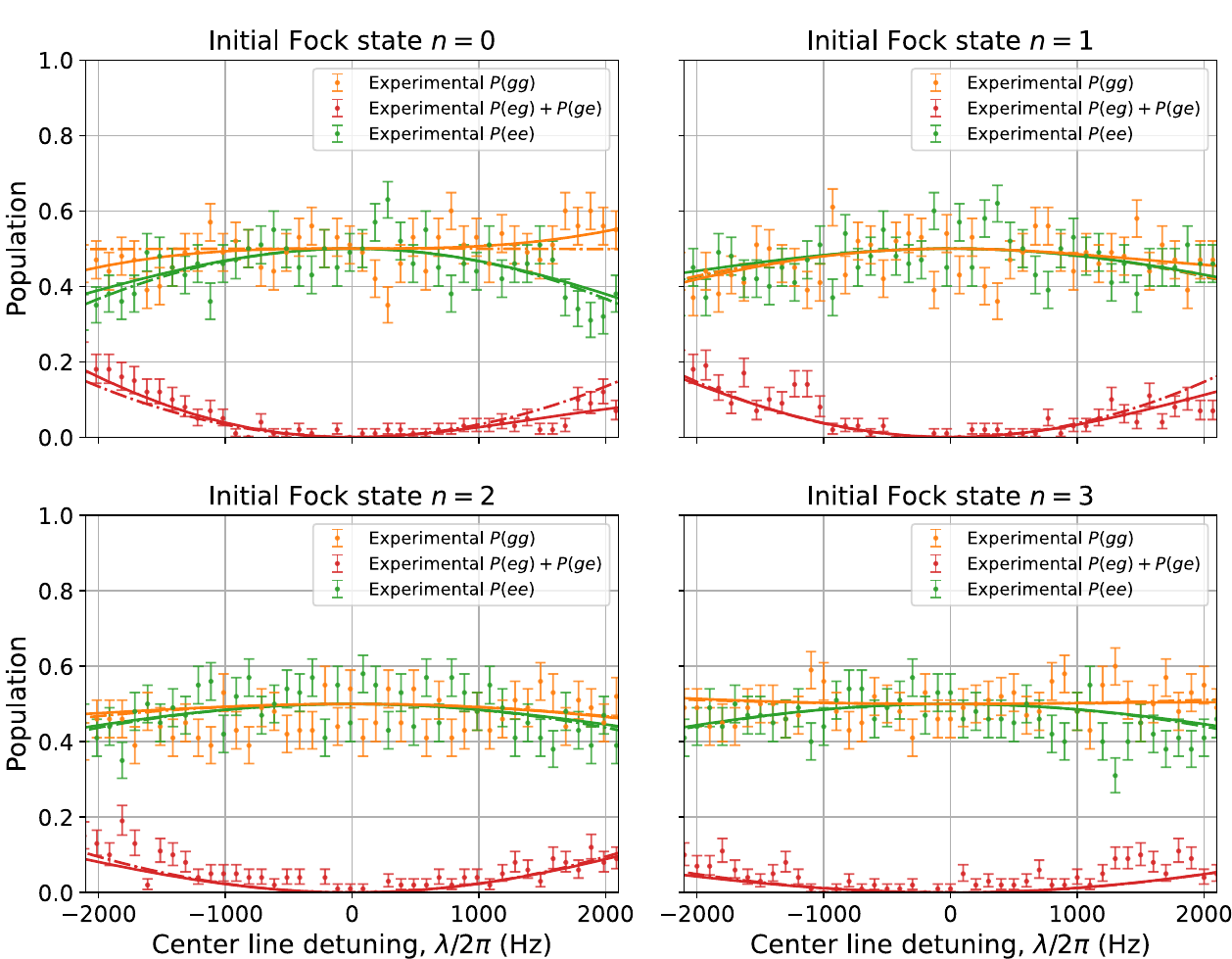}
    \caption{Experimental measurements of the populations obtained by application of a MS gate with $\epsilon=-2\pi\cdot 11$kHz to initial states $\ket{g,g,n}$ with $n=0,1,2,3$. The values obtained from the Magnus expansion are represented by the discontinuous lines, and the ones from numerical simulation by dashed-dotted lines.}
    \label{fig:experimental_populations_all}
\end{figure}

Using this setup, we can also study the behaviour of the populations after the application of a center line detuned gate in the experiment. In order to do this, we prepared $\ket{g,g,n}$ states with $n=0,1,2,3$, to which we then applied a single MS gate while scanning over the center line detuning. A comparison between the experimental results and the populations predicted by our model up to second order is shown in Fig.~\ref{fig:experimental_populations_all}.

\section{Conclusion \& outlook}
\label{sec:conclusion}

In this work we introduced a systematic analytical model for the characterisation of the effects that a center line detuning miscalibration has on the M\o lmer-S\o rensen gate. This model was obtained from a Magnus expansion where the center line detuning was considered as a perturbation to the ideal MS gate Hamiltonian. Using this approach we have shown how to predict the form of the final states obtained after application of the miscalibrated MS gate. Here, we performed the expansion up to first and second order in the center line detuning, by using a set of coefficients obtained from numerical integrations. It is then straightforward to understand the dependencies of relevant properties of the final states, such as relative phases, populations, fidelities, and purities, as functions characterised by these numerical coefficients. We then compared the prediction of these properties obtained from our theoretical model to results from numerical integration, finding only minor differences arising for higher center line detuning values due to influence of higher than second order terms, which we do not consider in our work. However, this discrepancy between model and numerical predictions appears for values of the center line detuning higher than those appearing during an experimental calibration of the MS gate. The value of the center line detuning miscalibration is typically only a fraction of the sideband detuning. 
Furthermore, we compared the predicted values of relative phases and populations from our model to values obtained from experimental measurements by systematically varying deliberately introduced center line detuning and find good agreement between the model predictions and the experimental results. The relationship between center line detuning, phase and phonon number has not been studied previously to our knowledge. For imperfectly cooled ions this may form a decoherence channel, as the thermal distribution of phonons is mapped to the phase of the applied gate. These results further validated our model, and confirmed the utility of our model for studying and improving experimental implementations of the MS gate. The predictions of the model for the populations are here limited by the order of the Magnus expansion. For example, the $P(gg)$ population for initial Fock state $n=0$ has a leading order term of third order, with the first two orders vanishing (see Fig.~\ref{fig:populations_comparison} and Fig.~\ref{fig:experimental_populations_all}). Thus a higher order expansion will be needed and can be realised based on our systematic treatment to accurately predict the behavior.

During the derivation of our model, we assumed for simplicity that the pulse shape of the laser used to implement the MS gate was constant. While this will not be exactly the case in the experimental implementation, the laser is usually shaped such that it has a relatively short (compared to the gate time) ramp-up time at the beginning, in which the laser intensity grows from zero up to its maximum value and, similarly, a short ramp-down time at the end of the gate, in which the intensity goes from this maximum value to zero. Therefore, during most of the gate time, the laser pulse is constant, justifying our approach. However, if one wanted to account for this effect, or even consider a general time-dependent laser pulse, this can be readily done by following a similar derivation as shown, but calculating the numerical coefficients appearing in the Magnus expansion by using the time-dependent form of the center line detuning shown in Eq.~\eqref{eq:time_dep_cl}. This could be specially useful when considering implementations of fast gates~\cite{Gaebler2016, schafer2018fast}, for which considering the laser pulse as constant might stop being a valid approximation.

Another consideration is that in the derivation of the MS gate Hamiltonian, we assumed the gate to be operating in first order Lamb-Dicke regime. However, since this regime is defined by $\eta \sqrt{n}\ll 1$, this is only valid if, given a value of the Lamb-Dicke parameter, the motional state of the ions has been cooled to a low enough value. This is the case for some experiments which implement an MS gate with thermal states of the order of $\bar{n}\approx 0.05$, while having a Lamb-Dicke parameter $\eta\approx 0.1$~\cite{schindler2013quantum}. However, outside of this regime, the appearance of higher order Lamb-Dicke terms could introduce an error and become the limiting factor of the gate performance. This could be the case of the previously mentioned fast MS gates, some of which rely on a higher value of $\eta$ in order increase the coupling to the sidebands. For the study of this case, a generalisation of our model considering such higher order Lamb-Dicke terms would be useful.

Finally, our model was derived by assuming that the center line detuning was the only miscalibration, but this will not be the case in a real implementation of a MS gate, where other parameters will differ from their ideal values, with some of these examples and their consequences discussed in Sec.~\ref{sec:ModelMS_ideal}. However, in the cases where the center line detuning miscalibration is relatively larger than for the other parameters, it will be the predominant effect. In this case one can detect that the outcomes of the miscalibrated gate agree with the results expected from the analytical model, and this information can be used to compensate for the miscalibration of the center line detuning. Furthermore, our theory could be extended to include other sources of miscalibrations, such as amplitude or gate time miscalibrations.

Overall, our model provides an in-depth understanding of the effects of a center line detuning in an implementation of the MS gate, which before had only been assessed by performing numerical calculations. This can then be used during the experimental calibration of the gate in order to identify and compensate the effect of center line detuning miscalibrations. Therefore, we believe that the method and the results presented here can help in designing and improving calibration routines for entangling gate operations. 

\section{Acknowledgements}

During the preparation of this manuscript, a related recent work~\cite{sutherland2021one} focusing on a systematic study of the impact of other imperfections on trapped-ion gate performance came to our attention. We thank M.~van Mourik and B.~Wilhelm for contributions to the experimental setup. We gratefully acknowledge support by the EU Quantum Technology Flagship grant AQTION under Grant Agreement number 820495, and by the US Army Research Office through Grant No. W911NF-14-1-010  and W911NF-21-1-0007.
We also acknowledge funding by the Austrian Science Fund (FWF), through the SFB BeyondC (FWF Project No. F7109), by the Austrian Research Promotion Agency (FFG) contract 872766, and by the IQI GmbH. MM acknowledges support by the ERC Starting Grant QNets Grant Number 804247. The research is also based upon work supported by the Office of the Director of National Intelligence (ODNI), Intelligence Advanced Research Projects Activity (IARPA), via the US Army Research Office Grant No. W911NF-16-1-0070. The views and conclusions contained herein are those of the authors and should not be interpreted as necessarily representing the official policies or endorsements, either expressed or implied, of the ODNI, IARPA, or the US Government. The US Government is authorised to reproduce and distribute reprints for Governmental purposes notwithstanding any copyright annotation thereon. Any opinions, findings, and conclusions or recommendations expressed in this material are those of the author(s) and do not necessarily reflect the view of the US Army Research Office.

\appendix

\section{Numerical coefficients}
\label{app:numerical_coefficients}

In the basis defined by the eigenstates of $S_y$, $\ket{+,+}$, $\ket{+,-}$, $\ket{-,+}$ and $\ket{-,-}$, where
\begin{equation}
    \ket{\pm}=\frac{1}{\sqrt{2}}(\ket{g}\pm i\ket{e}),
\end{equation}
the evolution operator of the ideal MS gate can be written as
\begin{equation}
\hat{U}_0(\tau)= 
\begin{pmatrix}
D[F(\tau)]e^{i G(\tau)} & 0 & 0 & 0 \\
 0 & 1 & 0 & 0 \\
 0 & 0 & 1 & 0 \\
 0 & 0 & 0 & D[-F(\tau)]e^{i G(\tau)} \\
\end{pmatrix},
\end{equation}
where $F(\tau) =\tilde\Omega (e^{i\tau} - 1 )$ and $G(\tau)=\tilde\Omega^2 (\tau - \sin\tau )$. From Eq.~\eqref{eq:M1} and Eq.~\eqref{eq:msm_magnus_evolution} we obtain
\begin{align}
    \hat{U}_{1,M}(\tau_g)\equiv\hat{U}_0&(\tau_g)M_1(\tau_g)\nonumber\\&=-i\tilde\lambda\int_{0}^{\tau_g} \hat{U}^\dagger_0(\tau'-\tau_g)\, S_z\, \hat{U}_0(\tau')d\tau'\label{eq:first_order_integral}.
\end{align}
Analysing the form of the integrand shows that its application couples a $\ket{+,+,n}$ state with $\ket{+,-,m}$ and $\ket{-,+,m}$, with corresponding coefficients $I^{++,n}_{+-,m}$ and $I^{++,n}_{+-,m}$, that is
\begin{equation}
\label{eq:++_first}
    \hat{U}_{1,M}(\tau_g)\ket{+,+,n}=-\tilde\lambda\sum_{m=0}I^{++,n}_{+-,m}\ket{+,-,m}+I^{++,n}_{-+,m}\ket{-,+,m}.
\end{equation}
Similarly, for the action on other $S_y$ basis states we obtain
\begin{align}
    &\hat{U}_{1,M}(\tau_g)\ket{-,-,n}=-\tilde\lambda\sum_{m=0}I^{--,n}_{+-,m}\ket{+,-,m}+I^{--,n}_{-+,m}\ket{-,+,m},\\
    &\hat{U}_{1,M}(\tau_g)\ket{+,-,n}=-\tilde\lambda\sum_{m=0}I^{+-,n}_{++,m}\ket{+,+,m}+I^{+-,n}_{--,m}\ket{-,-,m},\\
    &\hat{U}_{1,M}(\tau_g)\ket{-,+,n}=-\tilde\lambda\sum_{m=0}I^{-+,n}_{++,m}\ket{+,+,m}+I^{-+,n}_{--,m}\ket{-,-,m}\label{eq:-+_first}.
\end{align}
These coefficients are obtained from numerical integration of matrix elements as described in Eq.~\eqref{eq:first_order_integral}, and they can all be described in terms of $I^n_m\equiv I^{++,n}_{+-,m}$, 
\begin{align}
    &I^{++,n}_{+-,m}=I^{++,n}_{-+,m}=I^{+-,m}_{++,n}=I^{-+,m}_{++,n}=I^n_m,\\
    &I^{--,n}_{+-,m}=I^{--,n}_{-+,m}=I^{+-,m}_{--,n}=I^{-+,m}_{--,n}=(-1)^{n-m}I^n_m.
\end{align}
Therefore, it is enough to calculate the coefficients $I^n_m$, which have the form
\begin{equation}
    I^n_m=\frac{i}{2}\int_0^{\tau_g} e^{i G(\tau)}\bra{m}D[F(\tau)]\ket{n}d\tau,
\end{equation}
where for $m\geq n$
\begin{equation}
\label{eq:D_projection_great}
    \bra{m}D(\alpha)\ket{n}=\sqrt{\frac{m!}{n!}}\alpha^{m-n}e^{-\abs{\alpha}^2/2}\sum_{k=0}^n (-1)^k \binom{n}{k}\frac{\abs{\alpha}^{2k}}{(m-n+k)!},
\end{equation}
and for $m<n$
\begin{equation}
\label{eq:D_projection_less}
    \bra{m}D(\alpha)\ket{n}=\sqrt{\frac{m!}{n!}}(\alpha^*)^{n-m}e^{-\abs{\alpha}^2/2}\sum_{k=0}^m (-1)^{n-k} \binom{n}{k}\frac{\abs{\alpha}^{2(m-k)}}{(m-k)!}.
\end{equation}
By using these expressions, one can numerically calculate the $I^n_m$ coefficients for the first order Magnus expansion.

As for the second order coefficients, we have to work with Eq.~\eqref{eq:M2} and the second order terms of Eq.~\eqref{eq:msm_magnus_evolution} to obtain them. This requires to calculate the action of the following operators
\begin{align}
    &\hat{U}_0(\tau_g)M_2(\tau_g)=-\frac{\tilde\lambda^2}{2} \hat{U}_0(\tau_g)\nonumber\\
    &\times \int_{0}^{\tau_g}\int_{0}^{\tau_1}\left[\hat{U}^\dagger_0(\tau_1)\, S_z\, \hat{U}_0(\tau_1),\hat{U}^\dagger_0(\tau_2)\, S_z\, \hat{U}_0(\tau_2)\right]d\tau_2 d\tau_1,\label{eq:M2_integral_second}\\
    &\hat{U}_0(\tau_g)M_1(\tau_g)^2=\nonumber\\
    &-\tilde\lambda^2\int_{0}^{\tau_g}\int_{0}^{\tau_g} \hat{U}^\dagger_0(\tau_1-\tau_g)\, S_z\, \hat{U}_0(\tau_1)\hat{U}^\dagger_0(\tau_2)\, S_z\, \hat{U}_0(\tau_2)d\tau_2 d\tau_1,\label{eq:M1_integral_second}
\end{align}
where for convenience we define the combination of these operators as
\begin{equation}
    \hat{U}_{2,M}(\tau_g)=\hat{U}_0(\tau_g)\left(M_2(\tau_g)+\frac{M_1(\tau_g)^2}{2}\right)
\end{equation}
By close inspection of the previous integrands, one can see that they couple $\ket{+,+,n}$ and $\ket{-,-,n}$ to states of the form $\ket{+,+,m}$ and $\ket{-,-,m}$
\begin{align}
    \hat{U}_{2,M}(\tau_g)\ket{+,+,n}=-\tilde{\lambda}^2\sum_{m=0}J^{n}_{1,m}\ket{+,+,m}+J^{n}_{2,m}\ket{-,-,m},\label{eq:++_second}\\
    \hat{U}_{2,M}(\tau_g)\ket{-,-,n}=-\tilde{\lambda}^2\sum_{m=0}J^{n}_{2,m}\ket{+,+,m}+J^{n}_{1,m}\ket{-,-,m}.
\end{align}
Similarly, they couple $\ket{+,-,n}$ and $\ket{-,+,n}$ to $\ket{+,-,m}$ and $\ket{-,+,m}$
\begin{align}
    \hat{U}_{2,M}(\tau_g)\ket{+,-,n}=-\tilde{\lambda}^2\sum_{m=0}J^{n}_{3,m}(\ket{+,-,m}+\ket{-,+,m}),\\
    \hat{U}_{2,M}(\tau_g)\ket{-,+,n}=-\tilde{\lambda}^2\sum_{m=0}J^{n}_{3,m}(\ket{+,-,m}+\ket{-,+,m})\label{eq:-+_second},
\end{align}
where the coefficients, $J^{n}_{1,m}$, $J^{n}_{2,m}$, and $J^{n}_{3,m}$, can be calculated by numerically integrating Eq.~\eqref{eq:M2_integral_second} and Eq.~\eqref{eq:M1_integral_second} using Eq.~\eqref{eq:D_projection_great} and Eq.~\eqref{eq:D_projection_less}. Their expressions are

\begin{widetext}
 \begin{equation}
\begin{split}
J^{n}_{1,m} &=\frac{1}{4}   \int_{0}^{\tau_g}\int_{0}^{\tau_1}  e^{-i (G(\tau_1)- G(\tau_2)- G(\tau _g))}\braket{m|D(F(\tau_g)) D^{\dagger}(F(\tau_1))D(F(\tau_2))|n}d\tau_2 d\tau_1 
\\ &-  \frac{1}{4}   \int_{0}^{\tau_g}\int_{0}^{\tau_1} e^{i (G(\tau_1)- G(\tau_2)+G(\tau _g))}\braket{m| D(F(\tau_g)) D^{\dagger}(F(\tau_2))D(F(\tau_1))|n}d\tau_2 d\tau_1 \\ & + \frac{1}{4} \int_{0}^{\tau_g}\int_{0}^{\tau_g}  e^{i (G(\tau_2)-G(\tau_1-\tau _g))}\braket{m
 | D^\dagger(F(\tau_1-\tau _g ) ) D(F(\tau_2)) |n}d\tau_2 d\tau_1 
\end{split}
\end{equation}

\begin{equation}
\begin{split}
J^{n}_{2,m} &  = \frac{1}{4} \int_{0}^{\tau_g}\int_{0}^{\tau_1}  e^{-i (G(\tau_1)- G(\tau_2)- G(\tau _g))}\braket{m|D^{\dagger}(F(\tau_g)) D(F(\tau_1))D(F(\tau_2))|n}d\tau_2 d\tau_1 
\\ &-  \frac{1}{4}   \int_{0}^{\tau_g}\int_{0}^{\tau_1} e^{i (G(\tau_1)- G(\tau_2)+G(\tau _g))}\braket{m| D^{\dagger}(F(\tau_g)) D (F(\tau_2))D(F(\tau_1))|n}d\tau_2 d\tau_1 \\ & + \frac{1}{4} \int_{0}^{\tau_g}\int_{0}^{\tau_g}  e^{i (G(\tau_2)-G(\tau_1-\tau _g))}\braket{m
 | D(F(\tau_1-\tau _g ) ) D(F(\tau_2)) |n}d\tau_2 d\tau_1 
\end{split}
\end{equation}

\begin{equation}
\begin{split}
J^{n}_{3,m} &  = \frac{1}{8} \int_{0}^{\tau_g}\int_{0}^{\tau_1} e^{i (G(\tau_1)- G(\tau_2))} \braket{m|D^{\dagger }(F(\tau_1))D(F(\tau_2)) +  D(F(\tau_1))D^{\dagger}(F(\tau_2))|n}d\tau_2 d\tau_1 
\\ &-  \frac{1}{8}   \int_{0}^{\tau_g}\int_{0}^{\tau_1} e^{i (G(\tau_2)- G(\tau_1))} \braket{m|D^{\dagger}(F(\tau_2))D(F(\tau_1))-   D(F(\tau_2))D^{\dagger }(F(\tau_1))|n}d\tau_2 d\tau_1 \\ & + \frac{1}{8} \int_{0}^{\tau_g}\int_{0}^{\tau_g}  e^{i (G(\tau_1)- G(\tau_2))} \braket{m|D^{\dagger }(F(\tau_1))D(F(\tau_2))+  D(F(\tau_1))D^{\dagger }(F(\tau_2))|n}d\tau_2 d\tau_1 
\end{split}
\end{equation}
\end{widetext}

After obtaining the numerical coefficients $I^n_m$, $J^{n}_{1,m}$, $J^{n}_{2,m}$, and $J^{n}_{3,m}$, we can write the action of the center line detuned gate over the states $\ket{+,+,n}$, $\ket{+,-,n}$, $\ket{-,+,n}$ and $\ket{-,-,n}$. To consider the action over the states $\ket{g,g,n}$, $\ket{g,e,n}$, $\ket{e,g,n}$ and $\ket{e,e,n}$ it is enough to use
\begin{align}
    &\ket{g,g,n}=\frac{1}{2}\Big(\ket{+,+,n}+\ket{+,-,n}+\ket{-,+,n}+\ket{-,-,n}\Big),\\
    &\ket{g,e,n}=\frac{-i}{2}\Big(\ket{+,+,n}-\ket{+,-,n}+\ket{-,+,n}-\ket{-,-,n}\Big),\\
    &\ket{e,g,n}=\frac{-i}{2}\Big(\ket{+,+,n}+\ket{+,-,n}-\ket{-,+,n}-\ket{-,-,n}\Big),\\
    &\ket{e,e,n}=\frac{1}{2}\Big(-\ket{+,+,n}+\ket{+,-,n}+\ket{-,+,n}-\ket{-,-,n}\Big).
\end{align}

Using the previous relations and Eq.~\eqref{eq:++_first}-\eqref{eq:-+_first}, we obtain the first order correction to the states, $\ket{\psi^{(1)}_{\sigma,\sigma',n}}$, defined in Eq.~\eqref{eq:gg_first_correction}-\eqref{eq:ge_first_correction}. Similarly, Using Eq.~\eqref{eq:++_second}-\eqref{eq:-+_second}, we obtain the second order correction to the states, $\ket{\psi^{(2)}_{\sigma,\sigma',n}}$, defined in Eq.~\eqref{eq:gg_second_correction}-\eqref{eq:ge_second_correction}, where we introduced the following coefficients
\begin{align}
    &J^n_{+,m}=\frac{J^n_{1,m}+J^n_{2,m}+2J^n_{3,m}}{2},\\
    &J^n_{-,m}=\frac{J^n_{1,m}+J^n_{2,m}-2J^n_{3,m}}{2}.
\end{align}

Finally, from the form of the final state corrected up to second order when using the initial state $\ket{g,g,n}$, $\ket{\Psi^{(2)}_{g,g,n}}$, we can calculate the coefficients that are used in Sec.~\ref{sec:populations} and Sec.~\ref{sec:purity}:
\begin{align}
    &c_{g,g,n}=-\textrm{Re}\Big( J^n_{+,n}\Big)-\textrm{Im}\Big( J^n_{+,n}\Big)+ \sum_{m\ge 0} \abs{I^n_m+I^m_n}^2f^\textrm{even}_{n,m},\\
    &c_{e,e,n}=\textrm{Re}\Big( J^n_{-,n}\Big)-\textrm{Im}\Big( J^n_{-,n}\Big) + \sum_{m\neq n} \abs{I^n_m-I^m_n}^2f^\textrm{even}_{n,m},\\
    &c_{e,g,n}=\sum_{m\neq n} \abs{I^m_n}^2f^\textrm{odd}_{n,m}.
\end{align}
The corresponding coefficients when the initial state is in a different state of the computational basis can be calculated in a similar way.

\bibliographystyle{mio_apsrev4-1}
\bibliography{bibliography.bib}

\end{document}